\def\draftversion{false}

\RequirePackage{ifthen}
\ifthenelse{\equal{\draftversion}{true}}{
  \documentclass[aps,prb,10pt,galley,amsmath,amssymb,
                 superscriptaddress]{revtex4}
}{
  \documentclass[aps,prb,10pt,twocolumn,amsmath,amssymb,
                 longbibliography,superscriptaddress]{revtex4-1}
}

\usepackage{graphicx}
\usepackage[usenames,dvipsnames]{color} 
\usepackage{bm} 
\usepackage{hyperref} 
\hypersetup{		  
    colorlinks=true,
    linkcolor=blue,
    filecolor=blue,      
    urlcolor=blue,
    citecolor=blue
}
\usepackage{array}

\usepackage{soul}

\ifthenelse{\equal{\draftversion}{true}}{
  \marginparwidth 2.7in
  \marginparsep 0.5in
  \newcounter{comm} 
  \def\commnext{\stepcounter{comm}}
  \def\commtext{{\bf\color{blue}[\arabic{comm}]}}
  \def\commmar{{\bf\color{blue}[\arabic{comm}]}}
  \def\dvm#1{\commnext\marginpar{\small DV\commmar: #1}\commtext}
  \def\nvm#1{\commnext\marginpar{\small NV\commmar: #1}\commtext}
  \def\ism#1{\commnext\marginpar{\small IS\commmar: #1}\commtext}
  \def\mlab#1{\marginpar{\small\bf #1}}
  
  \def\parsedate #1:20#2#3#4#5#6#7#8\empty{#4#5/#6#7/20#2#3}
  \def\moddate{\expandafter\parsedate\pdffilemoddate{\jobname.tex}\empty}
  
  \newcommand{\eqlab}[1]{\Red{\hbox{\small\;\;[#1]}}\label{eq:#1}}
  \newcommand{\seclabel}[1]{\label{sec:#1}\Red{\small\;\;[Sec:~#1]}}
}{
  \def\dvm#1{}
  \def\nvm#1{}
  \def\ism#1{}
  \def\mlab#1{}
  
  \newcommand{\eqlab}[1]{\label{eq:#1}}
  \newcommand{\seclabel}[1]{\label{sec:#1}}
}

\newcommand{\sref}[1]{Sec.~\ref{sec:#1}}
\newcommand{\srefs}[2]{Secs.~\ref{sec:#1} and~\ref{sec:#2}}

\newcommand{\aref}[1]{Appendix~\ref{sec:#1}}
\newcommand{\fref}[1]{Fig.~\ref{fig:#1}}

\newcommand{\Fref}[1]{Figure~\ref{fig:#1}}
\newcommand{\nn}{\nonumber\\}
\newcommand{\beq}{\begin{equation}}
\newcommand{\eeq}{\end{equation}}
\newcommand{\bea}{\begin{eqnarray}}
\newcommand{\eea}{\end{eqnarray}}
\newcommand{\eq}[1]{Eq.~(\ref{eq:#1})}
\newcommand{\eqs}[2]{Eqs.~(\ref{eq:#1}) and (\ref{eq:#2})}
\newcommand{\eqo}[2]{Eqs.~(\ref{eq:#1}) or (\ref{eq:#2})}
\newcommand{\eqr}[2]{Eqs.~(\ref{eq:#1}-\ref{eq:#2})}
\newcommand{\Eq}[1]{Equation~(\ref{eq:#1})}

\newcommand{\ignore}[1]{}
\def\zt{\mathbb{Z}_2}

\def\k{{\bf k}}
\def\rr{{\bf r}}
\def\R{{\bf R}}
\def\G{{\bf G}}

\def\ket#1{\vert#1\rangle}
\def\bra#1{\langle#1\vert}
\def\ip#1#2{\langle#1\vert#2\rangle}
\def\me#1#2#3{\langle#1\vert#2\vert#3\rangle}

\def\O{\Omega}
\def\Ot{\widetilde{\Omega}}

\def\zo{{z\O}}
\def\dxy{{\Delta xy}}
\def\tzo{\theta_\zo}
\def\tdxy{\theta_\dxy}

\def\para{\partial_\alpha}

\def\parx{\partial_x}
\def\pary{\partial_y}

\def\Im{\textrm{Im}\,}

\def\bc{_\textrm{cut}}
\def\zc{z\bc}

\def\cS{{\cal S}}
\def\cC{{\cal C}}

\def\eee{\!=\!}

\def\znom{z_{\rm nom}}

\def\zt{Z_2}

\def\xhat{\hat{\bf x}}

\def\zhat{\hat{\bf z}}
\def\nhat{\hat{\bf n}}
\def\gp{g_\perp}
\def\gpar{g_\parallel}

\def\boc{_{\rm OC}}
\def\bbc{_{\rm BC}}
\def\buc{_{\rm UC}}

\def\cC{{\cal C}}

\def\ns#1#2{\{#1|c/#2\}}

\def\ee{^{\rm e}}
\def\oo{^{\rm o}}

\def\bmA{_{\mu A}}
\def\bmB{_{\mu B}}

\def\Nd{N_{\rm d}}
\def\Nu{N_{\rm u}}

\def\la{\langle\kern-2.0pt\langle}
\def\ra{\rangle\kern-2.0pt\rangle}

\begin{document}


\title{Axion coupling in the hybrid Wannier representation}

\author{Nicodemos Varnava}
\affiliation{
Department of Physics \& Astronomy, Rutgers University,
Piscataway, New Jersey 08854, USA}

\author{Ivo Souza}
\affiliation{Centro de F{\'i}sica de Materiales,
  Universidad del Pa{\'i}s Vasco (UPV/EHU), 20018 San Sebasti{\'a}n,
  Spain} \affiliation{Ikerbasque Foundation, 48013 Bilbao, Spain}

\author{David Vanderbilt}
\affiliation{
Department of Physics \& Astronomy, Rutgers University,
Piscataway, New Jersey 08854, USA}


\begin{abstract}

Many magnetic point-group symmetries induce a topological
classification on crystalline insulators, dividing them into
those that have a nonzero quantized Chern-Simons magnetoelectric
coupling (``axion-odd'' or ``topological''), and those that do not
(``axion-even'' or ``trivial'').  For time-reversal
or inversion symmetries, the resulting topological state is
usually denoted as a ``strong topological insulator'' or an
``axion insulator'' respectively, but many other symmetries can
also protect this ``axion $\zt$'' index.
Topological states are often insightfully characterized by
choosing one crystallographic direction of interest, and
inspecting the hybrid Wannier (or equivalently, the non-Abelian
Wilson-loop) band structure, considered as a function of the
two-dimensional Brillouin zone in the orthogonal directions.
Here, we systematically classify the axion-quantizing
symmetries, and explore the implications of such symmetries on
the Wannier band structure.  Conversely, we clarify the conditions
under which the axion $\zt$ index can be deduced from
the Wannier band structure.  In particular, we identify cases in
which a counting of Dirac touchings between Wannier bands,
or a calculation of the Chern number of certain Wannier
bands, or both, allows for a direct determination of the axion $\zt$ index.
We also discuss when such symmetries impose a ``flow'' on the
Wannier bands, such that they are always glued to higher and
lower bands by degeneracies somewhere in the projected Brillouin
zone, and the related question of when the corresponding surfaces
can remain gapped, thus exhibiting a half-quantized surface
anomalous Hall conductivity.  Our formal arguments are confirmed
and illustrated in the context of tight-binding models for
several paradigmatic axion-odd symmetries including time
reversal, inversion, simple mirror, and glide mirror symmetries.

\end{abstract}

\maketitle


\section{Introduction}
\seclabel{intro}

Recent progress in the classification of topological insulators (TIs)
has been dramatic.  Building on early work on the quantum anomalous
Hall (QAH) effect,\cite{haldane-prl88} the first phase in these
developments focused on systems with time-reversal (TR) symmetry.  In
this context, two-dimensional (2D) systems are known as spin-Hall
insulators, and three-dimensional (3D) insulators are classified as
weak or strong TIs.\cite{hasan-rmp10} While QAH insulators are
characterized by an integer topological invariant, the TR-protected
systems are described by $\zt$ invariants $\{0,1\}$ (or $\{+,-\}$)
corresponding to trivial and topological states respectively.  The
additional presence of inversion symmetry was shown
by Fu and Kane\cite{fu-prb07} to provide a
simple means of determining the strong $\zt$ index,
based on counting the numbers of odd-parity states at
high-symmetry points in the Brillouin zone (BZ).

The basic notion of the topological classification is the assignment
of two insulators to the same class if and only if they can be
connected by a continuous deformation of the Hamiltonian without gap
closure, while preserving a given set of symmetries.  We say that the
symmetries ``protect'' the topological classification.  At a formal
level, the list of protecting symmetries can include chiral and
particle-hole symmetries.\cite{schnyder-prb08,kitaev-aip09} However,
while relevant in the context of exotic phases such as
superconductivity, these are rarely of interest for the description of
ordinary insulators at the single-particle level.

In 2011, Fu introduced the concept of topological crystalline
insulators,\cite{fu-prl11} where the set of symmetries used for the
classification includes point symmetries, such as rotation or mirror
operations, in addition to TR.  The natural development of this trend
is to consider the space group -- or, for TR-broken systems, the
magnetic space group -- in setting up the topological classification.
Here there has been dramatic progress in recent years,
with new approaches
having emerged based on topological band theory\cite{jorrit-prx17,bradlyn-nat17}
and symmetry indicators,\cite{po-nc17} most recently including
magnetic groups.\cite{watanabe-sciadv18}
These have allowed for impressive progress
in clarifying the full variety
of possible topological states of crystalline insulators.

One tool that has proven especially useful for characterizing
topological states is the hybrid Wannier (HW) representation
(see Ref.~[\onlinecite{gresch-prb17}] for a recent overview).  Here
one selects one crystallographic direction, say $\zhat$, along which
to transform from $\k$ space to real space via the standard
construction of one-dimensional (1D) maximally localized Wannier
functions and their centers.  This is done independently at each $k_y$
in 2D, or at each $(k_x,k_y)$ in 3D. Thus, the HW functions are
localized in real space along $z$ but remain as extended Bloch-like
functions in the orthogonal direction(s).  The Wannier-center
positions can be obtained from a parallel transport analysis performed
independently for each $\k$-point string encircling the BZ in the
$k_z$ direction.  This kind of analysis also goes under the name of
the ``Wilson loop,''\cite{wilson-prd74,giles-prd81,wilczek-prl84},
the generalized non-Abelian Berry phases corresponding to the HW
centers are frequently referred to as ``Wilson-loop eigenvalues.''

For a $d$-dimensional insulator, the flow of these HW centers as a
function of wavevector in the orthogonal $(d-1)$-dimensional BZ, which
we shall refer to as the Wannier band
structure,\footnote{For a 3D insulator, the bulk band
  structure is defined in the 3D BZ, while both the surface energy
  band structure and the Wannier band structure are defined in a
  projected 2D BZ.}
often proves to be a very useful tool for determining its
topology.\cite{taherinejad-prb14,alexandradinata-prb14,
alexandradinata-prb16,varnava-prb18,bouhon-prb19,hwang-prb19}
For example, the integer Chern number of a 2D QAH system, the $\zt$
index of a 2D TR-invariant insulator, and the strong and weak indices
of a 3D TR-invariant insulator, are easily diagnosed via an inspection
of the flow of the HW centers.  They can also provide strong hints as
to the location in the BZ of the band inversion responsible for the
topological state, and to the flow of surface energy bands for
surfaces orthogonal to the wannierization
direction.

One interesting aspect of the 3D TR-invariant class of insulators is
that the $\zt$ index that distinguishes between strong-TI and trivial
(or weak-TI) systems also serves as an ``axion $\zt$
index.''\cite{qi-prb08,essin-prl09,wang-njp10} The axion coupling is
expressed in terms of a phase angle $\theta$, the ``axion angle,''
that constrains the possible values of the
anomalous Hall conductivity (AHC) on any insulating surface, as
measured relative to an outward unit normal, to take values
$(e^2/h)(N-\theta/2\pi)$ for integer $N$.  Since $\theta$ is mapped
into $-\theta$ under TR, and is only well defined modulo $2\pi$, the
presence of TR allows only two possible values of $\theta$, $0$ and $\pi$, which
correspond to the trivial and topological $\zt$ phases respectively.
This means that any insulating surface of a strong TI must exhibit a
half-integer surface AHC, as measured in units of $e^2/h$.  If all
facets of a crystallite are insulating and have the same value of
surface AHC (i.e., the same integer $N$), then the crystallite as a
whole behaves like a magnetoelectric material with an isotropic linear
magnetoelectric coupling of $\theta=(e^2/h)(\theta/2\pi-N)$.  In this sense,
$\theta$ is understood as describing a formal isotropic
magnetoelectric coupling.

In fact, this formal coupling is $\pi$ for ordinary
strong TIs such Bi$_2$Se$_3$ and its relatives,
even though they normally show no magnetoelectric effect.
This remains consistent with the previous considerations because
the surfaces are necessarily metallic unless TR is broken at the
surface.  The topologically protected Dirac cones on the surface
provide a canceling contribution, and the surface AHC vanishes
(as it must if TR is present).

On the other hand, it was shown that inversion symmetry also protects
the axion $\zt$ index.\cite{essin-prl09,hughes-prb11,turner-prb12}
When TR is absent and the system is
axion-odd, such systems are generally known as ``axion
insulators.''\cite{wan-prb11} A simple criterion was given by Turner
et al.\ for determining the axion $\zt$ index in the presence of
inversion symmetry,\cite{turner-prb12} generalizing the
parity-counting analysis of Fu and Kane\cite{fu-prb07} to the
TR-broken case.  Because the symmetry protecting the bulk topology
(i.e., constraining the values of $\theta$) is inversion instead of TR,
and because inversion is never a good symmetry at any surface, it is
much easier for the surface of an axion insulator to remain insulating.
In some
cases other symmetries may be present in addition and may force some
facets to be metallic, and there is always the possibility that
non-topological surface states will be present at the Fermi energy.
Nevertheless, if the goal is to find insulators that naturally display
a half-integer surface AHC response, then it appears that axion
insulators are much more promising. Unfortunately, physically
realized examples of 3D bulk crystals that behave as axion
insulators have been very difficult to find.  To date the closest to
a physical realization seems to the antiferromagnetic topological
insulator\cite{mong-prb10} MnBi$_2$Te$_4$, which has been the
subject of much recent interest.\cite{zhang-prl19,li-sa19,
chowdhury-njpcm19,otrokov-nat19,chen-nat19,yan-prm19}

Other symmetry operations also reverse the sign of $\theta$, and thus
support an axion $\zt$ classification.  These include simple mirrors
and glide mirrors, rotoinversions, and time-reversed rotations and
screws.  In general, an insulating material whose magnetic
space group contains such axion-odd operators has an axion $\zt$
index, and if that
index is nontrivial, the material is guaranteed to have a half-integer
QAH response on any insulating surface.

In this paper, we investigate the symmetry constraints imposed by the
presence of axion-odd symmetry operations on the Wannier band structure,
and more specifically, the additional constraints associated with
the axion-odd topological state.  Conversely, we show how
the axion $\zt$ index can often be deduced from an inspection
of the Wannier band structure. In many cases, this involves only 
a visual inspection of the Wannier band structure, possibly including
a counting of Dirac nodes between certain bands.  In other cases
it may require a calculation of the Chern number of a subset of
Wannier bands, or a more complicated computation of Berry fluxes on
truncated Wannier bands and Berry phases on the truncation boundaries.

The paper is organized as follows.
In \sref{HW-axion} we review the formal definition of the axion
coupling as an integral of the Chern-Simons (CS) three-form over the 3D
Brillouin zone (3DBZ).  We also introduce the HW representation and
the definitions of Berry potential and curvature on the Wannier bands.  We
then review the expression for the axion coupling in the HW
representation, which was originally derived only in the case of
isolated Wannier bands, and generalize these expressions to the case
where the
bands touch to form isolated groups, and to the case that the entire
Wannier band structure is connected.
In \sref{sym-cons} we classify the axion-odd symmetry operations
according to whether or not they reverse the $\zhat$ axis, and whether
or not they involve nonsymmorphic fractional translations along
$\zhat$.  For each of these classes, we discuss the simplifications
that occur because of cancellations in the expression for the axion
$\theta$, and clarify when and how the axion index can be
determined from an inspection of the Wannier band structure.
We also discuss which symmetries allow insulating vs.\ metallic surfaces 
in axion-odd insulators, and which allow gapped vs. gapless Wannier bands.
In \sref{cases} we discuss several common cases, such as the presence of TR or inversion by
itself, or the existence of a mirror or glide mirror symmetry.  We
illustrate each of these cases with explicit calculations for
a tight-binding model embodying the symmetry in question.
We summarize and conclude in \sref{summary}.
Finally, we also provide an Appendix in which the case of inversion
symmetry is treated by an alternative approach involving counting of
parity eigenvalues of the Wannier bands at high-symmetry $\k$ points,
leading to conclusions in agreement with the previous analysis.

\section{Hybrid Wannier representation and the axion coupling}
\seclabel{HW-axion}

\subsection{Axion coupling in the Bloch representation}
\seclabel{bloch}

We start from the usual expression for the axion coupling derived
from the second Chern number, a topological invariant
defined in 4D, through a dimensional reduction procedure.
One obtains the expression
\beq
\theta = -\frac{1}{4\pi}\int_{\rm BZ} d^3k \, \epsilon^{\alpha \beta \gamma} \text{Tr} \Big[ \mathcal{A}_\alpha \partial_\beta \mathcal{A}_\gamma -i\frac{2}{3} \mathcal{A}_\alpha \mathcal{A}_\beta \mathcal{A}_\gamma \Big] \, .
\eqlab{axioncoup}
\eeq
for the axion coupling of a 3D insulator in
the Bloch representation.\cite{qi-prb08,essin-prl09}
Here $\k=(k_x,k_y,k_z)$ runs over the 3DBZ and
$\mathcal{A}_\alpha$ is a shorthand notation for the non-Abelian Berry
connection $\mathcal{A}_{\alpha,nm}(\k) =
\bra{u_{n\k}}i\partial_{k_{\alpha}}\ket{u_{m\k}}$ obtained
from the cell-periodic Bloch functions $\ket{u_{n\k}}$.  The trace
in \eq{axioncoup} is over the occupied valence bands indexed by
$m$ or $n$.  The integrand of \eq{axioncoup} is known as the
CS three-form.

Being a (pseudo)scalar quantity characterizing the ground
state of a 3D insulator, one might expect the axion coupling
$\theta$ to be gauge-invariant, but this is not the case. In fact, a gauge
transformation, that is, a $\k$-dependent
unitary transformation $U_{mn}(\k)$ that
mixes occupied bands, can cause $\theta \rightarrow \theta + 2\pi N $
for some integer $N$. This implies that the only well defined part of
the bulk axion coupling lives in an interval of $2\pi$.

Physically, the axion coupling
describes an isotropic contribution to the formal magnetoelectric (ME)
tensor\cite{essin-prl09+e,hasan-rmp10}
$\alpha_{ij} = ( \partial P_i/\partial B_j )_{\bf E} = ( \partial
M_{j}/\partial E_{i} )_{\bf B}$. In other words,
an orbital magnetic field will induce a parallel polarization, or
equivalently, an electric field will induce a parallel orbital
magnetization, with a constant of proportionality given by
\beq
\alpha_{\rm CS}
= \frac{e^2}{h}\frac{\theta}{2\pi} \, .
\eqlab{alphaiso}
\eeq
We described this as a ``formal'' ME response because it
manifests itself through the appearance of a surface AHC on any
insulating surface facet that is given by
\beq
\sigma^{\text{surf}}_{\text{AHC}} =
    -\frac{e^2}{h}\frac{\theta}{2\pi} \ \text{mod} \ e^2/h \, .
\eqlab{surfahc}
\eeq
In this exact relation, the ambiguity modulo $2\pi$ in
$\theta$ is consistent with a freedom to prepare insulating
surfaces with values of $\sigma^{\text{surf}}_{\text{AHC}}$ differing
by the quantum $e^2/h$, e.g., by changing the Chern number of some
surface bands, or of adding or deleting a surface layer with a nonzero
Chern number.\cite{qi-prb08,essin-prl09+e,coh-prb11,rauch-prb18,
  vanderbilt-book18} If all surfaces adopt the same branch choice --
i.e., the same value of $\sigma^{\text{surf}}_{\text{AHC}}$ -- then
the sample as a whole exhibits a true magnetoelectric response of
$-\alpha_{\rm CS}$, where the quantized part of the response has
been absorbed into the branch choice for $\alpha_{\rm CS}$.

This phenomenon is a higher-dimensional analog
of the modern theory of electric polarization,\cite{kingsmith-prb93}
where the $2\pi$ ambiguity of the Berry phase reflects the inability
to define the bulk polarization,
or to predict the bound charge density of an insulating
surface, except modulo a quantum.

\subsection{The hybrid Wannier representation}
\seclabel{HW}

For a given crystalline 3D insulator, we choose a
reciprocal-lattice direction along which the HW
representation will be constructed, and orient the
Cartesian axes such that the corresponding real-space direction is
$z$.  Given a gauge for the Bloch states $\ket{\psi_{n\k k_z}}$, the
corresponding HW states are expressed as
\beq
\ket{h_{ln\k}}= \frac{1}{2\pi} \int^{\pi/c}_{-\pi/c} dk_z e^{-ik_z l c}
  \ket{\psi_{n\k k_z}} \,.
\eqlab{h-def}
\eeq
Here $\k=(k_x,k_y)$ is the wavevector in the perpendicular plane,
i.e., in the projected two-dimensional Brillouin zone (2DBZ);
$c=2\pi/b$ is the lattice constant along $z$, with $b$ the
magnitude of the shortest reciprocal lattice vector along $z$;
$l$ is an index that runs over unit
cells along $z$; and $n$ runs over the $J$ occupied bands in the
insulator, which is the same as the number of HW functions at each
$\k$ in one vertical unit cell.

Henceforth the multiband gauge of the Bloch wavefunctions is
always taken such that the $\ket{h_{ln\k}}$ are maximally
localized along $z$.\cite{marzari-prb97}
The HW wavefunctions $h_{ln\k}(\rr) =\ip{\rr}{h_{ln\k}}$ are
thus Bloch-like and cell-periodic\footnote{By ``cell-periodic''
we mean that the $e^{i\k\cdot\rr}$ phase has been factored out,
so that $h_{ln\k}(\rr)=h_{ln\k}(\rr)+\R$ for any in-plane lattice
vector $\R$.}  in the in-plane directions, while at each
$\k$ they are the maximally localized Wannier functions of the
effective 1D Hamiltonian $H_\k$ at that $\k$.  Their centers
\beq
z_{ln}(\k)=\me{h_{ln\k}}{z}{h_{ln\k}}
\eqlab{z-def}
\eeq
form the Wannier ``bands'' (or ``sheets'').  These are
periodic in real space along $z$, $z_{ln}(\k)=z_{0n}(\k)+lc$, as well
as periodic in the in-plane reciprocal space,
$z_{ln}(\k)=z_{ln}(\k+\G)$, where $\G$ is an in-plane reciprocal
lattice vector.  We shall frequently drop the explicit $\k$ dependence
in the following.  As noted earlier, this is essentially the same
construction as that of the non-Abelian Wilson loop.

The Berry connection is defined in the HW representation as
\beq
A^\alpha_{ln,l'n'}=\me{h_{ln}}{i\para}{h_{l'n'}},
\eqlab{Adef}
\eeq
where $\partial_\alpha=\partial/\partial_{k_\alpha}$ with
$\alpha=\{x,y\}$.  As a reminder, $l$ and $l'$ run over unit cells in
the $z$ direction, and $n$ and $n'$ label Wannier bands within the
unit cell.  Since there are $J$ occupied bands in our insulator, $n$
and $n'$ run from 1 to $J$.  We shall also need to make use of the
Wannier-band-diagonal Berry curvature
\bea
\O_{ln,ln} &=& \parx A^y_{ln,ln}- \pary A^x_{ln,ln} \nn
             &=& -2\,\Im\ip{\parx h_{ln}}{\pary h_{ln}} \,.
\eqlab{O-def}
\eea
Periodicity implies that
\beq
A^\alpha_{ln,l'n'}=A^\alpha_{0n,(l'-l)n'}
\eqlab{A-per}
\eeq
and therefore $\O_{ln,ln}=\O_{0n,0n}$.

Next, we discuss the gauge dependence of the quantities introduced
above, as first presented in Ref.~[\onlinecite{taherinejad-prl15}].
The Wannier bands are predetermined by the maximal localization procedure,
so except in the case of degeneracy between Wannier bands,%
\footnote{In the case that $N$ Wannier bands are degenerate in some region
  of the 2DBZ, the present arguments can be generalized by considering
  a multiband gauge transformation involving a $\k$-dependent
  $N\!\times\!N$ unitary mixing among the degenerate Wannier bands; after
  tracing over any physical contribution over the degenerate bands,
  the same conclusions follow.}
the most general gauge transformation that we need to consider
is a Wannier-band-dependent phase twist
\beq
\ket{\tilde{h}_{ln}}=e^{-i\beta_{ln}(\k)}\,\ket{h_{ln}} \,.
\eqlab{gauge-h}
\eeq
This leads to new Berry connections
\beq
\tilde{A}^\alpha_{ln,l'n'}=
e^{-i(\beta_{l'n'}-\beta_{ln})}\,(A^\alpha_{ln,l'n'}
  +\delta_{ln,l'n'}\para\beta_{ln}) \,.
\eqlab{gauge-A}
\eeq
The Berry curvature, on the other hand, is gauge invariant,
which follows from the first line of \eq{O-def} using
$\parx\pary\beta=\pary\parx\beta$.

We shall also have reason to define the ($\k$-dependent) quantity
\beq
\Gamma_{ln,l'n'}=i\,(z_{l'n'}-z_{ln})\,A^x_{ln,l'n'}\,A^y_{l'n',ln}
\eqlab{Gam-def}
\eeq
which is fully gauge invariant, i.e.,
\beq
\tilde{\Gamma}_{ln,l'n'} = \Gamma_{ln,l'n'} \,.
\eqlab{gauge-Gamma}
\eeq
To see this, note that the $(z_{l'n'}-z_{ln})$
prefactor in \eq{Gam-def} insures that only
off-diagonal elements of $A^\alpha$ contribute, and that the product
of $A^x_{ln,l'n'}$ and $A^y_{l'n',ln}$ leads to a cancellation of the
phase factors in \eq{gauge-A}.

\begin{figure}
\centering\includegraphics[width=\columnwidth]{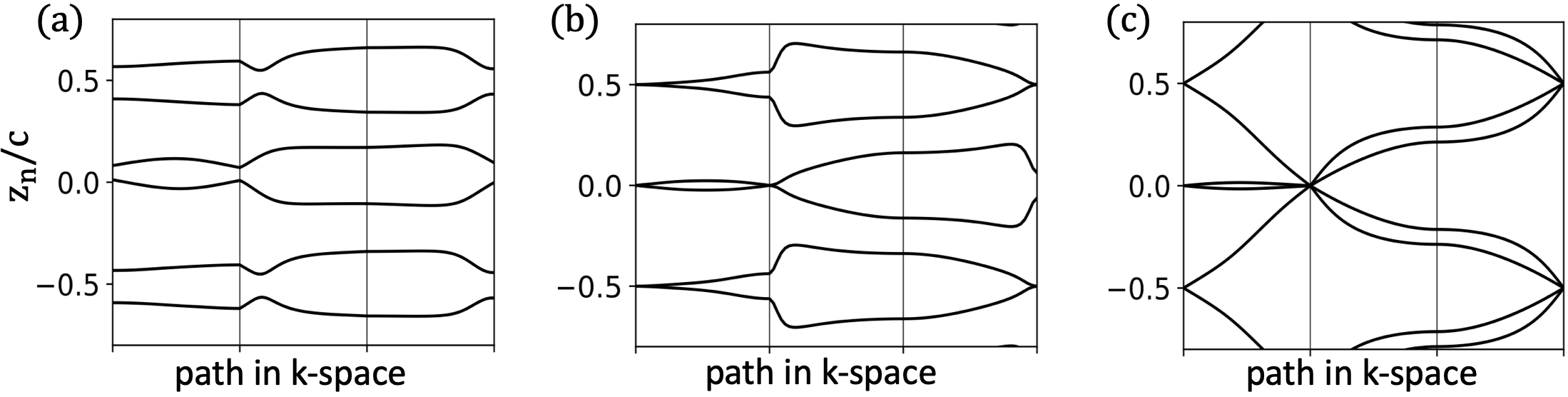}
\caption{Illustrative sketches of possible hybrid Wannier (i.e.,
  Wilson loop) band structures for a model with four occupied bands.
  Hybrid Wannier (HW) centers $z_{nl}(\k)$ repeat along the vertical direction
  with period $c$.  The horizontal axis represents some path
  connecting high-symmetry points in the 2D Brillouin zone.  (a) Four
  isolated Wannier bands. (b) Two connected groups of Wannier bands.
  (c) Fully connected Wannier bands.}
\label{fig:hwbands}
\end{figure}

Finally we discuss the connectivity (sometimes called the ``flow'')
of the Wannier bands.
We restrict ourselves to the case that all bulk Chern indices
are zero. We refer to the vanishing of the Chern number in the
$x$-$y$ plane as the ``in-plane Chern constraint.''  The
vanishing of the other two Chern indices guarantees that
each Wannier band returns to itself (not to higher or lower partners)
as one traverses the 2DBZ by a reciprocal lattice vector.  Of course, it
also returns to itself on any closed loop that does not wind by
a reciprocal lattice vector, since all such loops are contractible.%
\footnote{The fact that the bulk is globally insulating is also
  important. In a Weyl semimetal, for example, traversing a loop
  that winds around the projection of a Weyl point in the 2DBZ
  pumps the Berry phase by $2\pi$, so the Wannier bands cannot return
  to themselves around such a loop.}
Importantly,
these considerations allow us to label the Wannier bands globally by
integers that we take as increasing along $\zhat$.

A \textit{gap} is said to exist between a pair of adjacent HW bands
if these are not connected by degeneracies anywhere in the 2DBZ.  A Wannier
band is said to be \textit{isolated} if a gap exists above and below
it.
A \textit{connected group}
of Wannier bands is a set of adjacent bands
that are connected by degeneracies, but that are separated by gaps above and
below the group.  The HW sheet structure as a whole is said to be
\textit{connected} if there are no gaps; otherwise it is
\textit{disconnected} and is composed of $M$ internally
connected groups separated by $M$ gaps per unit cell along $z$.  These
features are illustrated in \fref{hwbands}.

\subsection{Disconnected Wannier band structure}
\seclabel{z-disconnected}

\subsubsection{Isolated bands}
\seclabel{iso}

Let us begin with the simplest case, in which all Wannier bands are
isolated, so that $z_{ln}$ is a smooth function of $\k$ for each $n$.
For this case, which is illustrated in \fref{hwbands}(a), the authors
of Refs.~[\onlinecite{taherinejad-prl15}] and [\onlinecite{olsen-prb17}]
showed that the CS axion coupling $\theta$ can be expressed in
the HW representation as
\beq
\theta=\tzo+\tdxy \,.
\eqlab{th-a}
\eeq
The first term
\beq
\tzo=-\frac{1}{c}\int d^2k \sum_n z_{0n}\O_{0n,0n}
\eqlab{tzo-pre}
\eeq
is a kind of ``Berry curvature dipole'' term, where $z_{0n}$ is given
by \eq{z-def} and the summation index $n$ runs over Wannier bands in the
home unit cell ($l\eee0$).  The second term in \eq{th-a} can be
written as
\beq
\tdxy= -\frac{1}{c}\int d^2k \sum_{n}\sum_{l'n'} \Gamma_{0n,l'n'}
\eqlab{tdxy}
\eeq
where $\Gamma_{ln,l'n'}$ was defined in \eq{Gam-def}.
We have chosen to fix index $l\eee0$, but the
result is unchanged if the index 0 is replaced by arbitrary
$l$ in \eq{tdxy}, since $\Gamma$ obeys the same kind of translational
invariance as in \eq{A-per}, i.e.,
\beq
\Gamma_{ln,l'n'}=\Gamma_{0n,(l'-l)n'} \,.
\eqlab{G-per}
\eeq

Henceforth we simplify the notation by establishing the
convention that the absence of a cell index $l$ implies $l\eee0$,
i.e., $z_n$ is shorthand for $z_{0n}$, $\O_{nn}$ is shorthand for
$\O_{0n,0n}$, etc.  Then \eq{tzo-pre} becomes just
\beq
\tzo=-\frac{1}{c}\int d^2k \sum_n z_n\O_{nn} \,.
\eqlab{tzo}
\eeq
Similarly, $\ket{h_n}$ shall refer to a HW function $\ket{h_{0n}}$ in the
home unit cell $l\eee0$.

While all the quantities in \eqs{tdxy}{tzo} are gauge invariant, so
that there is no ambiguity on this account, there is an important
ambiguity of a different kind in $\tzo$.  It comes from the freedom to
choose the set of Wannier bands assigned to the home unit cell; this
affects $\tzo$ because of the appearance of $z_n$ in \eq{tzo}. By
contrast, only differences of $z$ values appear in \eq{tdxy}, so this
term is fully unambiguous.

To understand the ambiguity in $\tzo$, recall that the labels
$n=\{1,2,...,J\}$ simply count the bands in ascending order
within the home unit cell.  (As a reminder, we are considering here the
case that all Wannier bands are isolated.) But how do we choose the unit
cell? Which band shall we label as $n\eee1$?  One way to think about
this is that the choice of home unit cell corresponds to the choice of
one of the $J$ gaps as the
``primary'' one, such that the counting starts with
the Wannier band just above this gap.
A different choice of primary gap
just has the effect of shifting some subset of the Wannier bands along $z$
by distance $c$.  For each shifted band $n$, $\tzo$ in \eq{tzo}
gets shifted by
\beq
\Delta\tzo=-\int d^2k\,\O_{nn} = -2\pi C_n\,,
\eqlab{Dt-C}
\eeq
where $C_n$ is the Chern number computed over this Wannier band.
Since $C_n$ is necessarily an integer, this means that $\tzo$, and
also the total $\theta$ as given by \eq{th-a}, is
only well defined modulo $2\pi$.  Actually, such an ambiguity is
expected, since we know on other grounds that $\theta$ is
only well defined modulo $2\pi$, so this is not problematic.
However, it is still something that we have to anticipate and deal
with in the analysis that follows.

To clarify why $\tdxy$ is unaffected by the choice of unit cell,
it is instructive to rewrite \eq{tdxy} as
\beq
\tdxy= -\frac{1}{Nc}\int d^2k \sum_{\mu\mu'} \Gamma_{\mu\mu'} \,,
\eqlab{tdxy-r}
\eeq
where we have introduced a condensed index notation $\mu\eee(ln)$, the
sums over $\mu$ and $\mu'$ both run over all Wannier bands in a large
system of $N$ unit cells, and
$\Gamma_{\mu\mu'}=i(z_{\mu'}-z_\mu)A^x_\mu A^y_{\mu'}$.  In other
words, $\tdxy$ is just the sum of all $\Gamma$ elements taken per unit
cell along $\zhat$.  With this perspective, it is obvious that the
$ln$ labeling of the Wannier bands is irrelevant for
$\tdxy$.  By contrast, it is impossible to write an expression similar
to \eq{tdxy-r} for $\tzo$, since the appearance of $z_\mu$ itself,
rather than the difference $z_{\mu'}-z_\mu$, would render the
average over $N$ cells ill-defined.

\subsubsection{Composite groups of bands}
\seclabel{comp}

We now consider the case that at least some of the bands form
internally connected composite groups, but the Wannier band
structure as a whole remains disconnected.  An example of a system of
this type is shown in \fref{hwbands}(b), where there are $M\eee2$
connected groups, each consisting of a pair of bands joined by a nodal
point.  We do not expect such nodal points to appear generically,
since accidental degeneracies have codimension three, and thus require
fine tuning.  On the other hand, such nodal degeneracies may
sometimes be induced by symmetry, often occurring at the
time-reversal invariant momenta (TRIM),
or at other high-symmetry points or lines, in the 2DBZ.  When such
degeneracies are present, the evaluation of \eq{th-a} becomes
problematic because $\O_{nn}$ can diverge in the vicinity of the
degeneracies between Wannier bands.  In the case of a Dirac node, defined
as an isolated nodal touching with linear dispersion of $z_n(\k)$
close to the node, $\Omega_{nn}$ has a delta-function singularity at
the node.  To solve this problem, we can go over to a formulation in
terms of a gauge-covariant treatment of the Berry curvature within
each group.

To see this, let the $a$'th group ($a=1,...,M$) be composed of ${\cal
M}_a$ Wannier bands, and combine terms such that the contribution of
group $a$ is taken to be
\beq
\theta_a= -\frac{1}{c}\int d^2k \left(
  \sum_{n\in a} z_n\O_{nn} +\sum_{n,n'\in a} \Gamma_{nn'}
  \right) \,.
\eqlab{th-cg}
\eeq
As a reminder, $z_n$, $\O_{nn'}$, and $\Gamma_{nn'}$ refer to contributions
coming from the home cell $l\eee0$ (and $l'\eee0$) only.
Then the total Chern-Simons coupling is
\beq
\theta=\tzo'+\tdxy'
\eqlab{th-b}
\eeq
where
\beq
\tzo'= \sum_a \theta_a
\eqlab{thta}
\eeq
is the sum of \eq{th-cg} over groups $a$, and
\beq
\tdxy'= -\frac{1}{c}\int d^2k \sum_n{\sum_{l'n'}}' \Gamma_{0n,l'n'}
\eqlab{tdxyp}
\eeq
is identical to $\tdxy$ in \eq{tdxy} except that the prime on
the sum indicates the omission of all terms with Wannier bands
$l'n'$ belonging to the same group as $0n$.
The sum over groups in \eq{thta} counts isolated bands by
treating them as groups with
${\cal M}_a\eee1$, and \eq{th-a} with \eqs{tdxy}{tzo}
are recovered if all bands are isolated.

But now the quantity inside the parentheses in \eq{th-cg}
can be simplified by writing it as
\beq
\theta_a= -\frac{1}{c}\int d^2k \sum_{n\in a} z_n\Ot_{nn}
\eqlab{th-cg-b}
\eeq
where
\beq
\Ot_{nn} = \O_{nn}-i\sum_{m\in a}
        (A^x_{nm}A^y_{mn} -A^y_{nm}A^x_{mn})
\eqlab{Ot-def}
\eeq
is the diagonal element of a gauge-covariant Berry curvature matrix.
Substituting \eq{Ot-def} into \eq{th-cg-b} easily demonstrates
the equivalence of these expressions.
The Chern number of a connected group of bands can then be expressed in
terms of the gauge-covariant Berry curvature as
\beq
C_a=\frac{1}{2\pi} \int d^2k\sum_{n\in a}\Ot_{nn} \,.
\eqlab{Cgroup}
\eeq

The advantage of this formulation is that $\Ot_{nn}$ remains a smooth
and divergence-free function of
$\k$ in the vicinity of degeneracies between bands in the group.  This
is a well-known feature of the gauge-covariant Berry curvature, and
can be seen by expressing it as
\beq
\Ot_{nn} = -2\,\Im\ip{\tilde{\partial}_x h_{n}}{\tilde{\partial}_y h_{n}} \,,
\eqlab{Ot-Q}
\eeq
where
$\tilde{\partial}_\alpha \ket{ h_{n}}=Q_a \partial_\alpha \ket{ h_{n}}$
is the gauge-covariant derivative of the HW state and
$Q_a=1- \sum_{n\in a}\ket{h_n}\bra{h_n}$ is the projector onto all
states other than the Wannier bands in group $a$ in the home unit cell.
(The philosophy is similar to that used in defining the gauge-covariant
derivative
$\tilde{\partial}_\alpha\ket{u_{n\k}}=Q_{n\k} \partial_\alpha\ket{u_{n\k}}$
of the Bloch energy eigenstate $\ket{u_{n\k}}$,
where $Q_{n\k}=1-\sum_{m\ne n}\ket{u_{m\k}}\bra{u_{m\k}}$,
except that here we
work with the spectrum of $z$, not that of $H$.) The projector $Q_a$
eliminates from $\tilde{\partial}_\alpha \ket{ h_{n}}$ the
divergences that would be present in $\partial_\alpha \ket{ h_{n}}$
arising from mixing between Wannier bands inside the same group.

In summary, in this part we have argued that
\eqr{th-b}{th-cg-b} form a robust set of equations that
can be used to evaluate the CS coupling even in the
presence of composite groups of internally-degenerate Wannier bands.

In case there is a doubt about the correctness of this expression,
we can consider the application of a small symmetry-lowering
perturbation $\lambda V$ that gaps out the degeneracies between
bands within the group.  In this case we know \eq{th-a} is correct,
and we can argue that it is equivalent to \eq{th-b},
and then take the limit $\lambda\rightarrow0$.  Since \eq{th-b}
is insensitive to degeneracies within the group, we can
conclude that its evaluation at $\lambda\eee0$ provides the
correct CS coupling.

\subsection{Introduction of a cutting surface}
\seclabel{cutsurf}

\begin{figure*}
\centering\includegraphics[width=5.4in]{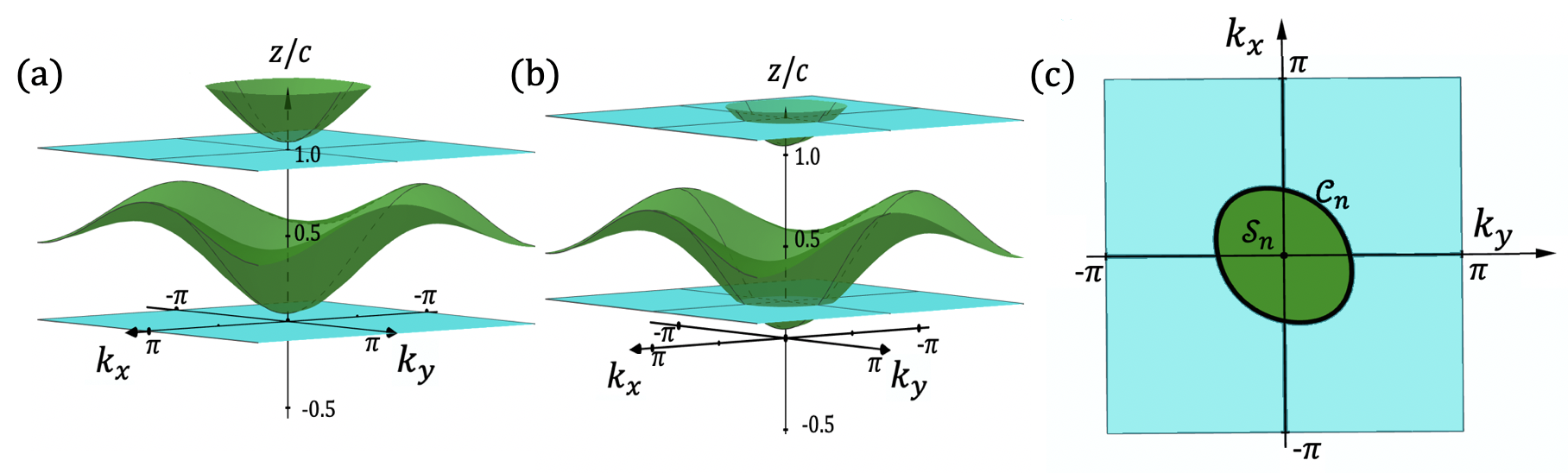}
\caption{(a) An isolated Wannier band (green) lying entirely inside the
  home unit cell (cutting surfaces, blue, at $z\eee0$ and $z\eee
  c$). (b) The same Wannier band, now represented by two disconnected
  pieces in the home unit cell, after the cutting surface has been
  raised.  (c) The region $\cS_{n}$ where the Wannier band falls below the
  new cutting surface, and its boundary $\cC_{n}$.}
\label{fig:cut}
\end{figure*}

When it comes time to consider the case of a fully connected Wannier band
structure, it will not be possible to assign Wannier bands to the home unit
cell without cutting through the Wannier bands in some way.  To prepare for
this, we begin by returning to the case of isolated Wannier bands as in
\sref{iso}, and we define a ``cutting surface'' $\zc(\k)$ that is
smooth and periodic in $\k$, such that all Wannier bands with
$\zc(\k)<z_{ln}(\k)<\zc(\k)+c$ are assigned to the home cell
$l\!=\!0$.  As a reference, if the cutting surface lies entirely
inside the primary gap, as illustrated in \fref{cut}(a), then the
definition of the home unit cell is the same as it was in \sref{z-disconnected}.

Instead, let $\zc(\k)$ be increased such that it cuts through one or
more of the Wannier bands, as illustrated for a single Wannier band in
\fref{cut}(b). This has the effect that $z_n$ is shifted upwards by
$c$ for all bands lying below the new cut.  Following an earlier
argument, the contribution of Wannier bands lying entirely below the cut is
changed by $2\pi$ times a Chern integer, making no change to $\theta$
modulo $2\pi$.  As for band $n$ pierced by the cut, we define $\cS_n$
to be the region of the 2D plane for which $z_n$ lies below the cut,
and $\cC_n$ is its boundary, i.e., the intersection with the cut, as
shown in \fref{cut}(c). The reassignment of the HW states inside
$\cS_n$ shifts $z_n$ upwards by $c$, so that the term $\tzo$ in
\eq{tzo} changes by an amount
\beq
\Delta\tzo^{(n)}=-\int_{\cS_n} \!\! d^2k\,\O_{nn} = -\phi\bc^{(n)}
\eqlab{Dt-frac}
\eeq
where
\beq
\phi\bc^{(n)} = \oint_{\cC_n} {\bf A}_{nn}\cdot d\k
\eqlab{Berry-def}
\eeq
is the Berry phase evaluated on $\cC_n$.
In general, the boundary $\cC_n$ could be multiply connected, in which
case the sum over loop Berry phases is implied in \eq{Dt-frac}.  The
total change in $\tzo$ is then
\beq
\Delta\tzo=-\phi\bc=-\sum_n\phi\bc^{(n)} \,,
\eqlab{Dt-frac-tot}
\eeq
where $\phi\bc$ is the total Berry phase from all Wannier bands that
intersect $z\bc(\k)$.
Of course, this quantity is only well defined modulo $2\pi$, but
this is not an issue since the axion coupling has the same
indeterminacy.

On the other hand, the expression for $\tdxy$ in \eq{tdxy} is
unchanged, since no matter the labeling, all pairs of Wannier bands at the
same $\k$ eventually enter the sum in \eq{tdxy} in the same way as
before.  The overall change in \eq{th-a} is then just
$\Delta\theta=-\phi\bc$.  To correct for this, we just have to add
back a piece to cancel \eq{Dt-frac-tot}, and we arrive at
\beq
\theta=\tzo+\tdxy+\phi\bc \,.
\eqlab{th-c}
\eeq
As a reminder, $\tzo$ is still evaluated as in \eq{tzo}, but
now with the band label $n$ running from 1 to $J$
independently at each $\k$ beginning with the first band above the
cutting surface, and the Berry-phase term $\phi\bc$ accounts for the
contributions of the loops of intersection of the cutting surface with
the bands.  \Eq{th-c} is one of the principal results of the present
work.

\subsection{Connected Wannier band structure}
\seclabel{connected}

The result in \eq{th-c} was derived for the case that all bands are
isolated, but we now wish to consider a fully connected Wannier band
structure, in which no bands are isolated and no gaps occur.

\subsubsection{Degeneracy regions}
\seclabel{zOmega-gen}

As a first step, we revise \eq{th-c} for the case of composite
groups
considered in \sref{comp}.  In particular, we consider the
case that the cutting surface $\zc(\k)$ cuts through one or more
of the connected bands within a group (but avoiding degeneracies).
Unfortunately, we cannot simply cut through a connected group that
is being treated using the gauge-covariant formulation of
\sref{comp}, because $\phi\bc$ is defined as the Berry phase on
an individual Wannier band; this is equal to the integral of the
$\O_{nn}$ over the enclosed area, but not that of $\Ot_{nn}$.

\begin{figure}[b]
\centering\includegraphics[width=0.5\columnwidth]{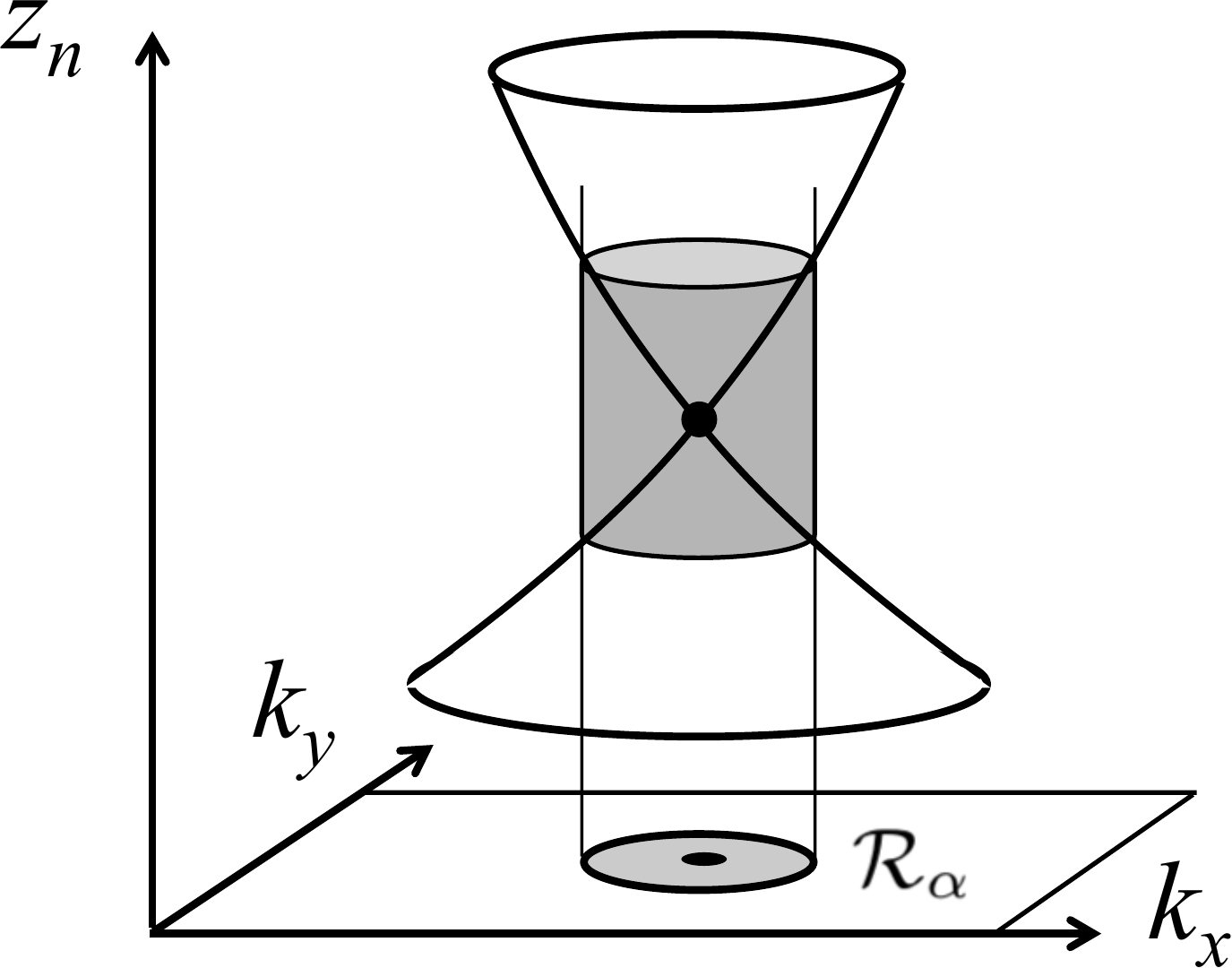}
\caption{Sketch of degeneracy region (shaded) associated with two
  Wannier bands $z_n(\k)$ near a point node, and its projection onto the
  disk-shaped region ${\cal R}_\alpha$ in the 2DBZ (bottom).}
\label{fig:dr}
\end{figure}

To remedy this problem, we can adopt a more restrictive treatment of
degeneracies, as follows.  Wherever there is a degeneracy between
Wannier bands, we identify a degeneracy region (DR) surrounding the
degeneracy.  The $\alpha$'th DR consists of a set of ${\cal M}_\alpha$
adjacent bands (${\cal M}_\alpha\ge2$) that are involved in the
degeneracy, and a small region ${\cal R}_\alpha$ surrounding the
degeneracy in the 2DBZ, as illustrated in \fref{dr}.
If a cutting surface $\zc(\k)$ is present,
we shall insist that it be chosen to avoid all the DRs, so that each DR lies
entirely inside the home unit cell.  We then collect together the
terms in \eqr{th-a}{tdxy} that only involve bands inside the DR
to get a contribution
\bea
\theta^{\rm DR}_\alpha 
&=& -\frac{1}{c}\int_{{\cal R}_\alpha} d^2k
  \left( \sum_{n\in\alpha} z_n\O_{nn}+\sum_{n,n'\in\alpha}
         \Gamma_{nn'} \right) \nn
&=& -\frac{1}{c}\int_{{\cal R}_\alpha} d^2k \sum_{n\in\alpha} z_n\Ot_{nn} \,.
\eqlab{tDR}
\eea
Here we have followed the same approach leading to \eq{th-cg}, but
now restricting the $\k$ integral only to the region ${\cal R}_\alpha$,
and the sum to run only over the ${\cal M}_\alpha$ bands involved in the
degeneracy.

We then write the total Chern-Simons axion coupling in \eq{th-c} as
\beq
\theta=\tzo''+\tdxy''+\phi\bc \,,
\eqlab{th-d}
\eeq
where
\beq
\tzo''= -\frac{1}{c}\int d^2k {\sum_n}'' z_n\O_{nn}
 +\sum_\alpha\theta^{\rm DR}_\alpha
\eqlab{tht}
\eeq
and
\beq
\tdxy''= -\frac{1}{c}\int d^2k \sum_n{\sum_{l'n'}}'' \Gamma_{0n,l'n'}
\,.
\eqlab{tdxypp}
\eeq
The double prime on the first sum in \eq{tht} indicates that all terms coming
from within a DR are to be omitted, and in \eq{tdxypp} it excludes
terms where both bands lie in the same DR in the same cell;
these omissions are compensated by the second term in \eq{tht}.

\Eq{th-d} is the desired
formula, which remains robust in the presence of degeneracies,
including in the connected case in which no gaps are present.
It provides the formal solution to the problem of expressing the
axion coupling $\theta$ in the HW representation, even in the case of
fully connected Wannier bands, and is one of our principal results.

While \eq{th-d} might be somewhat awkward to implement in practice,
involving as it does the choice of some DRs that need
to be treated differently while integrating over the 2DBZ and
summing over bands, we shall mainly be interested below
in cases in which some symmetry is present that quantizes $\theta$
to 0 or $\pi$. 
In such cases, we shall insist on choosing the cutting surface and
the DRs in such a way as to respect those symmetries, so that the
same kinds of symmetry arguments used for the HW sheet contributions
can also be used for the DR contributions.  Thus, in cases where
the $\tzo$ contribution would vanish in the absence of DRs,
$\tzo''$ vanishes in their presence as well.  In other cases, we shall argue
shortly that one can take a limit in which the size of the DRs goes to
zero.  Either way, an explicit calculation of the contribution of
a DR can typically be avoided. In fact, it often happens that
only the last term $\phi\bc$ survives in
\eq{th-d}, so that the axion coupling is given just by the Berry
phase on the cutting loop (or the total Berry phase in the case of
multiple loops).

Of course, whenever the Wannier band structure is actually disconnected, it
is simpler to return to \eq{th-b}, where no cutting surface is needed
and the bands can be indexed by counting from above some chosen gap.
Again, symmetry will often allow us to decide the value of $\theta$
based on rather general features of the Wannier band structure in this case
as well.

\subsubsection{Shrinking the degeneracy regions}
\seclabel{shrink}

Up to this point, we have avoided specifying the nature of the
degeneracies between Wannier bands, which in general could occur at
point nodes, or along lines, or even over a plane spanning the 2DBZ,
depending on the type of symmetries present.  Henceforth
we will focus on point touchings of two or more Wannier bands, commenting
only occasionally on the case of higher-dimensional degeneracies.
Then, from a formal point of view, we can shrink the size of the
$\alpha$'th DR surrounding a point node to a disk of some small radius
$\epsilon$ in the 2DBZ.
This may be problematic computationally,
since the immediate vicinity of the DR may become difficult
to treat without the gauge-covariant formulation, but as a formal
manipulation it is permissible.  Furthermore, we can argue that
the contribution $\theta_\alpha^{\rm DR}$ vanishes in the limit
that $\epsilon\rightarrow0$, since $\Ot_{nn}$ remains finite as the
degeneracy is approached (since it only ``sees'' Wannier bands outside
the degenerate group), and the area of the disk goes to zero.

Formally speaking, then, we can simply neglect the contributions from
the DRs in the small-DR limit.
This will prove useful in analyzing the contributions to $\theta$
in the presence of certain symmetries, as we shall see.

\section{Symmetry considerations}
\seclabel{sym-cons}

\subsection{Axion-odd insulators}
\seclabel{gen-axion}

For the remainder of the manuscript we restrict ourselves to
insulating systems with symmetries that protect the quantization of
$\theta$ to 0 or $\pi$.  This symmetry could be time reversal (TR), in
which case a spinful system with $\theta=\pi$ is usually denoted as a
strong topological insulator (TI).  It also could be inversion $I$, in
which case the system is usually called an axion
insulator.  However, many other symmetries can quantize $\theta$,
such that $\theta/\pi=0$ or 1 defines an ``axion $\zt$ index.''
We use the term ``axion-odd insulators'' to refer
to systems in which the nontrivial $\zt$ index is protected by one of
these symmetries, with TR-protected strong TIs and inversion-protected axion
insulators as special cases.

In this section we survey the symmetries that can quantize the axion
coupling, and describe their consequences for the Wannier band structure.
Moreover, we show that in many cases it is possible to
determine the axion $\zt$ index without recourse to the expressions
given in \sref{HW-axion}.  In particular, it is often enough just to
have a knowledge of some elementary features of the Wannier bands, such as
the type and number of touchings between bands, or the total Chern
numbers of certain bands or band groups.

To decide whether $\theta$ is quantized to 0 or $\pi$ by a set
of crystal symmetries -- i.e., whether the axion $\zt$ index
is protected -- we can just look at whether there are any elements
in the magnetic point group that reverse the sign of $\theta$.
We shall call these the ``axion-odd'' symmetries, and they
are comprised of the proper rotations composed with TR and the
improper rotations not composed with TR. If one or more of these
symmetries is present in the magnetic point group, then
$\theta$ is quantized to be 0 or $\pi$, i.e., the $\zt$ index exists.

In such a case, we can argue as follows that the
$\tdxy$ term in \eq{tdxy} must vanish.  Like the $\tzo$ term,
$\tdxy$ has its sign reversed by any axion-odd symmetry operation.
On the other hand, unlike $\tzo$, $\tdxy$ has no quantum of
ambiguity, as we saw at the end of \sref{iso}.  As a result, $\tdxy$
is immediately forced to vanish in the presence of such a symmetry,
while $\tzo$ is not.  The restricted sums $\tdxy'$ and $\tdxy''$ in
\eq{th-b} and (\ref{eq:th-d}), respectively, will vanish as well,
since the assignment of bands to composite groups automatically
respects the symmetry, and we chose the shapes and locations of the
degeneracy regions to respect the symmetry as well.  For this
reason, we ignore the $\tdxy$ terms in the considerations that
follow, and the remaining question is whether the $\tzo$
term, taken together with $\phi\bc$ if a
cutting surface is present, yields 0 or $\pi$.

In many cases the magnetic point group may contain several axion-odd
symmetry operations, but any one of them is enough to quantize
$\theta$.  Therefore, we shall just consider each kind of axion-odd
point symmetry in turn, and study its consequences, while keeping in
mind that other symmetries may exist as well.

Let $g$ be the axion-odd magnetic point symmetry in question.
For this $g$, we choose our Cartesian frame such that the 
direction $\zhat$ is either invariant or reversed by $g$,
and then we carry out the wannierization along the $\zhat$ direction.
In some cases this choice can be made in more than one way; for
example, for a mirror, $\zhat$ could be chosen to lie in, or normal
to, the mirror plane.  In such cases more than one avenue of
investigation may be available, with one possibly being more
advantageous, but for now we just assume that some such choice
has been made.

In this chosen frame, the possibilities for the spatial part of the
point-group operator can be enumerated as follows.  If $\zhat$ is not
reversed, it is the identity $E$; a proper rotation $C_n$ by $2\pi/n$
about the $\zhat$ axis; or a reflection $M_d$ across a plane
containing the $\zhat$ axis.  If $\zhat$ is reversed, it is the inversion
$I$; a mirror $M_z$ across the $x$-$y$ plane; an improper rotation
$S_n=M_zC_n$ for $n=\{3,4,6\}$; or a two-fold rotation $\bar{C}_2$
about an axis lying in the $x$-$y$ plane.  We attach a prime to denote
composition with TR, so that $E'$ represents TR itself.  This
establishes our notations for the symmetry operations.  We also
decompose $g=\gpar\,\gp$, where $\gpar$ is $E$ or $M_z$, and $\gp$ is
the in-plane part of the operator, including the TR component if
present. For orientation, we list in Table~\ref{tab:g} the
axion-odd point-group symmetry operations $g$ along with their decomposition
into $g_\parallel$ and $g_\perp$.

\begin{table}
\caption{\label{tab:g}Axion-odd point-group symmetries, and their
  decomposition as $g=g_\parallel g_\perp$.}
\begin{ruledtabular}
\begin{tabular}{p{2.3cm}p{2.3cm}p{2.3cm}p{2.3cm}}
& $g$ & $g_\parallel$ & $g_\perp$ \\
\hline
\multicolumn{4}{l}{Operations reversing $\zhat$} \\
& $M_z$        & $M_z$ & $E$         \\
& $I$          & $M_z$ & $C_2$       \\
& $S_{3,4,6}$  & $M_z$ & $C_{3,4,6}$ \\
& $\bar{C}_2'$ & $M_z$ & $M_d'$      \\
\multicolumn{4}{l}{Operations preserving $\zhat$} \\
& $E'$           & $E$   & $E'$ \\
& $C'_{2,3,4,6}$ & $E$   & $C'_{2,3,4,6}$ \\
& $M_d$          & $E$   & $M_d$ \\
\end{tabular}
\end{ruledtabular}
\end{table}

Keeping in mind that we are restricting ourselves to the case that
$g$ is axion-odd, we can distinguish three general cases:
\begin{enumerate}
\item Operation $g$ reverses $\zhat$.  We assume a choice of
$z$-origin such that the corresponding space-group operation has no
fractional translation along $\zhat$.%
\footnote{If the original choice of origin is such that the operation
  takes $z\rightarrow z_0-z$, then we reset the origin at the
  invariant point $z_0/2$.}
\item Operation $g$ preserves $\zhat$, and the corresponding
space-group operation has no fractional translation along $\zhat$.
\item Operation $g$ preserves $\zhat$, and the corresponding space-group
operation involves a fractional translation $c\zhat/m$.
\end{enumerate}
The first case corresponds to $g_\parallel\eee M_z$, while the second
and third pertain to $g_\parallel\eee E$.  In all cases there may also
be a fractional (nonsymmorphic)
in-plane translation ${\boldsymbol\tau}_\perp$ in the space-group
operation associated with $g$; these play a role in determining
when and where point nodes of degeneracy may occur between neighboring
Wannier bands, but will not concern us here.  If point nodes do occur,
these will typically be located at $z\eee0$ or $z\eee c/2$ for
$g_\parallel\eee M_z$, and at general locations for
$g_\parallel\eee E$.

We denote the axion-odd space-group operations as being 
either $z$-reversing, $z$-preserving, or $z$-nonsymmorphic
according to the cases enumerated above.
In the following subsections we shall consider each of these
cases in turn.

Before proceeding, we list the
transformation rules for the Wannier bands and their Berry
curvatures. Under an operation
$\{g\vert c\zhat/m+{\boldsymbol\tau}_\perp\}$ of the space group, a
Wannier band transforms as~\cite{taherinejad-prb14}
\beq
z_{n'}(\pm g_\perp\k)=g_\parallel z_n(\k)+c\zhat/m\,,
\eqlab{z-transf}
\eeq
where the minus sign applies when $g_\perp$ includes TR.  The
Berry curvature is a pseudovector pointing along $\zhat$, and as a
result it transforms as
\beq
\O_{n'n'}(\pm g_\perp\k)=
\begin{cases}
  \O_{nn}(\k),& \text{$g$ reverses $\zhat$},\\
  -\O_{nn}(\k),& \text{$g$ preserves $\zhat$},
\end{cases}
\eqlab{O-transf}
\eeq
and the same rules apply to the gauge-covariant Berry
curvature $\Ot_{nn}$.

\subsection{Determination of the quantized axion coupling}
\seclabel{axion-det}

\subsubsection{Symmetry operation reverses $\zhat$}
\seclabel{rev}

For the operations that reverse $\zhat$, listed in the first
four rows of Table~\ref{tab:g}, it follows from
\eq{z-transf} that each HW center at $(z,\k)$ in band $n$ has a
partner at $(-z,\pm g_\perp\k)$ in some band $n'$ (again, the minus
sign applies when $g_\perp$ includes TR).  Moreover,
according to \eq{O-transf} the Berry curvatures of the Wannier bands are
identical at these locations.
At least in a simple situation such as that in \sref{iso}, it
therefore appears at first sight that all contributions to the $\tzo$
term given by \eq{tzo} will cancel in pairs when summing over all
Wannier bands and integrating over the 2DBZ, because
of the sign reversal of $z$.  However, it is not always possible to
choose the home unit cell in such a way that the cancellation in
\eq{tzo} is complete.

\paragraph{Disconnected case.}
\seclabel{rev-disconnected}
%
To see this, first consider the case of a disconnected Wannier band structure.
As discussed earlier, we arrange the Wannier bands into a set of
internally connected
groups that are isolated from one another by gaps, and we work
with \eqs{thta}{th-cg-b}, which were formulated for this case.
Clearly the arrangement of HW groups has to respect the
$z\leftrightarrow-z$ symmetry, so we can proceed as follows.

\newcommand{\OC}{origin-centered}
\newcommand{\BC}{boundary-centered}
\newcommand{\UC}{uncentered}

We organize the internally connected band groups into three
collections as follows.
First, we see whether there is a connected group centered at
$z\eee0$, and if so, we call it the ``origin-centered''
group.  In general, this can be done by listing all connected groups that
pass through $z\eee0$. If the number of such groups is odd,
the central one is the \OC\ group, and if not, there is no \OC\ group.
A similar construction at $z\eee c/2$ allows us to identify the
``boundary-centered''
group, if there is one.  We then see
whether there are any remaining HW groups lying between the \OC\ and
\BC\ groups
(or between the corresponding gaps if these groups are
absent).  If so, we collect these, together with their
$z\!\rightarrow\!-z$ symmetry partners lying
below the \OC\ group, into an ``uncentered''
collection of band groups.  This leads to a unique procedure
for defining the home unit cell as composed of the union of the
\OC\ group, the \BC\ group, and the \UC\ collection.
\Fref{zflip}(a) shows an example in which there are origin-centered
and boundary-centered groups composed of two Wannier bands each, and
an uncentered collection accounting for two more Wannier
bands, for a system with six occupied valence bands.

\begin{figure}
\centering\includegraphics[width=\columnwidth]{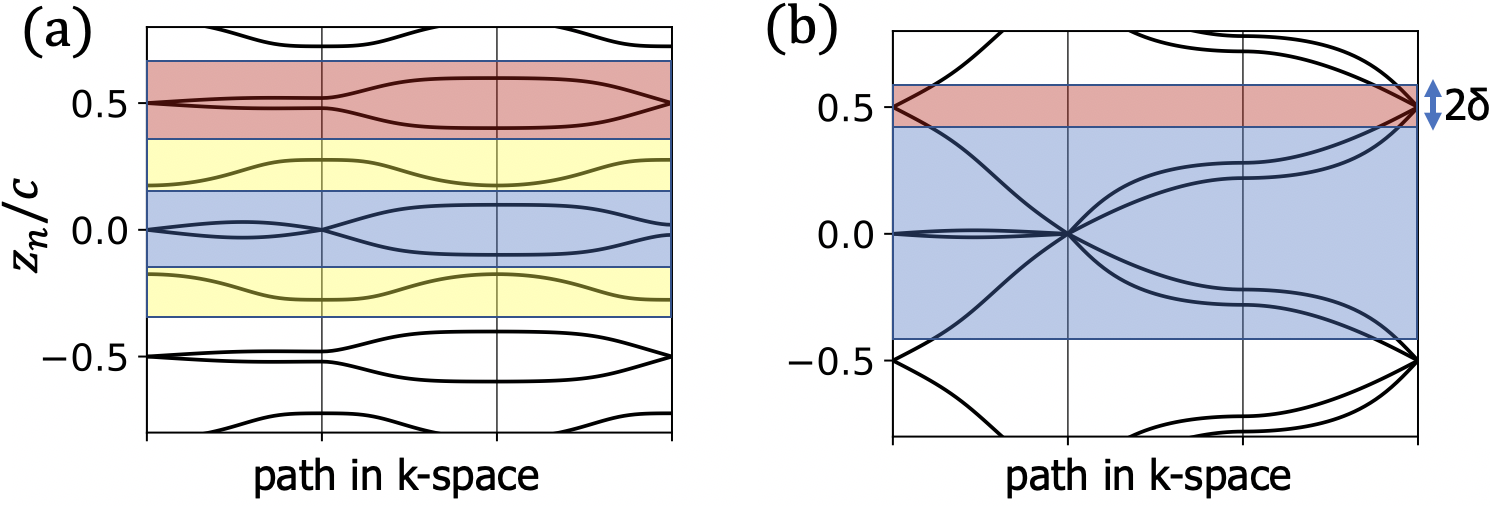}
\caption{Sketches of hybrid Wannier band structures along a path
  in k-space that captures all the degeneracies
  of a system with an axion-odd symmetry that reverses $\zhat$.
  (a) Disconnected Wannier band structure,
  decomposed into an origin-centered group (blue shading),
  a boundary-centered group (red shading), and an uncentered
  collection (yellow shading). The axion index depends on whether
  the Wannier bands in the red region have an odd Chern index, which
  in turn is determined by a counting of Dirac nodes;
  here $\theta\eee0$ because the number of nodes in that region
  is even.
  (b) Connected Wannier band structure, decomposed into a thin
  slice centered at $z\eee c/2$ (red shading) and the remainder
  (blue shading).  Here $\theta=\pi$ since there is an
  odd number of Dirac cones in the red region.\\
  }
\label{fig:zflip}
\end{figure}

We can also obtain the contributions of these three collections to
various physical properties.
Let $C\boc$, $C\bbc$, and $C\buc$ denote the total in-plane Chern
numbers of the \OC, \BC, and \UC\ collections respectively,
as computed from \eq{Cgroup}.
We can also obtain the contribution of each collection to $\tzo'$
in \eq{thta}.  We argued earlier that $\tdxy'=0$ in the presence
of any axion-odd symmetry, so we denote the $\tzo'$ contributions as simply
$\theta\boc$, $\theta\bbc$, and $\theta\buc$ respectively,
where we have also dropped the prime for brevity.  Then the
total axion coupling is just
\beq
\theta=\theta\boc+\theta\bbc+\theta\buc \,.
\eqlab{CCCC}
\eeq

Now the contribution $\theta\boc$ from the origin-centered group takes
the form given in \eq{th-cg-b}, and this clearly vanishes in view of
the cancellations between contributions at $z$ and $-z$ inside this
Wannier band group (recall that $\Ot_{nn}$ has the same sign in the canceling
pair).  Similar arguments imply that $\theta\buc$
also vanishes, since the groups below and above $z\eee0$
cancel in pairs.  However, there is no such cancellation for the Wannier
bands in the \BC\ group, because their symmetry-mapped partners are
centered about $z\eee-c/2$ and are outside the chosen home unit cell
(see \fref{zflip}(a), where the home cell consists of the union of the
four shaded regions).  Instead, each Wannier band at $(z,\k)$ in the \BC\
group has a partner in the same group at $(c-z,\pm\gp\k)$.  Again
$\Ot_{nn}$ is identical at both locations, so when inserting $z_n$
into \eq{th-cg-b}, we make no mistake if we treat both as located at
the average location $z=c/2$.  But then \eq{th-cg-b} becomes
\beq
\theta\bbc=-\frac{1}{2}\int d^2k \sum_{n\in {\rm BC}} \Ot_{nn}
= -\pi \, C\bbc 
\eqlab{half}
\eeq
where \eq{Cgroup} has been used.
Since this is the only contribution, we have that $\theta=\pi$
if and only if the Chern index of the \BC\ group is an odd integer.

Now recall that we assumed that all bulk Chern numbers vanish in
our material, so that
$C\bbc+C\boc+C\buc=0$.  Clearly $C\buc$ is even, since the
uncentered collection is always composed of groups that contribute
in pairs.  Thus, $C\bbc$ is odd if and only if $C\boc$ is odd. That is,
the existence of
an odd-Chern \BC\ group also implies the existence of an odd-Chern
\OC\ group. We are thus guaranteed to get the same result using the
\BC\ or \OC\ group in \eq{half}. The same conclusion follows from the freedom
to shift the origin by $c/2$ along $z$.

\paragraph{Counting of Dirac nodes}
\seclabel{rev-node-count}

In some cases it may be possible to deduce the Chern number
of a boundary- or origin-centered group by inspecting the nodal
touching of Wannier bands.  For example, suppose there are only two
Wannier bands in the boundary-centered group, and that the locus of
degeneracies between these bands consists only of some number
$K$ of Dirac point nodes.\footnote{To be a Dirac node, the dispersion
  $z_n(\k)$ must be linear for all in-plane $\hat{\k}$ directions.
  This excludes quadratic and other higher-order nodal points
  from our discussion here.}
In view of \eq{z-transf}, these are typically found at $z\eee c/2$
in the Wannier direction, and at locations obeying $\pm g_\perp\k=\k$
in the 2DBZ.  For example, they may occur at some of the
four projected TRIM (PTRIM) for $g\eee E'$, or at other
high-symmetry points or even at generic positions in the 2DBZ for
other symmetries.

Because we are considering symmetries that preserve the Berry
curvature while interchanging $z$ and $c-z$, and thus interchanging the
two bands, we know that each Wannier band carries the same
Berry flux $\Phi_0$.  Now, if the symmetry is weakly broken in such
a way as to gap the Dirac nodes, then an additional Berry flux of
$\pm\pi$ is transferred to each Wannier band at each of the
nodes. However, this has to result in an integer Chern number, so
we conclude that $\Phi_0$ must be $0$ or $\pi$ (mod $2\pi$) if the
number $K$ of Dirac nodes is even or odd, respectively. Recalling
\eq{half},
this means that the axion $\zt$ index is odd only if $K$ is
odd.  The same analysis can be applied to the origin-centered group.

This argument easily generalizes to the case that the
number ${\cal M}\bbc$ of Wannier bands in the group is even, still
assuming that the only touchings are Dirac-like.
That is, the Chern number is odd if and only if the total
number $K_\textrm{tot}$ of Dirac nodes is odd.  In this counting procedure,
a node involving an $L$-fold degeneracy is counted as $L/2$
Dirac nodes.

If the number ${\cal M}\bbc$ of Wannier bands in the group is
odd, then we have to treat the central band separately.
We let $j=({\cal M}\bbc+1)/2$ be the index of this band,
and assume that it is regular enough to have a well defined
Chern number $C_j$.
Each nodal touching involving this band will also involve
$L/2$ pairs of neighbors from among the remaining $({\cal M}\bbc-1)$
bands. These can be treated as before, providing $L/2$
Dirac nodes at one touching location, and a total number
$K_\textrm{tot}$ when summed over the 2DBZ.
The conclusion is that the total
Chern index of the group, and thus the axion $\zt$ index, is
odd if and only if $C_j+K_\textrm{tot}$ is odd.  This case is
not quite as convenient, because it requires the evaluation
of the Chern index of the central band, in addition to a simple
counting of Dirac nodes.

In this analysis, we have assumed that the locus of contact between
Wannier bands consists only of point Dirac nodes.
If cases arise involving high-order
(e.g., quadratic) point nodes, or nodal lines or regions, then the
analysis given above would need to be reconsidered.

\paragraph{Connected case.}
\seclabel{rev-connected}

Essentially identical results involving the counting of Dirac
nodes, and possibly the calculation of the Chern index of a central
band, can be derived for the case of a connected Wannier band
structure.  To do so, we first establish a ``nominal cell
boundary'' $\znom(\k)$ located near $z\eee-c/2$ as follows.  Let $N$
be the number of Wannier bands passing through the point
$(-c/2,\bar{\Gamma})$.  If $N$ is odd, let $\znom(\k)$ be identified
with the central one of these bands; if $N$ is nonzero and even, let
it be the average of the central two bands; and if $N\eee0$, let it be
the average of the next lowest and next highest bands about $z\eee-c/2$.
Then $\znom(\k)$ is a surface centered at $-c/2$ that respects the
symmetries of the system and passes through the point
$(-c/2,\bar{\Gamma})$.

If $N$ is even, $\znom(\k)$ lies between two Wannier bands.
We assumed a connected band structure, so even if $N\eee0$
these bands must touch at one or more nodes, which lie
on the surface $\znom(\k)$ by construction.
If $N$ is odd, then $\znom(\k)$ is identified with a HW band, which
again touches with the next higher band by assumption.  In either
case, let the cutting surface be
\beq
z\bc(\k)=\znom(\k)+\delta \,,
\eqlab{shift}
\eeq
where $\delta>0$ is a small vertical shift.
With this convention, the unit cell consists of
all portions of the Wannier bands lying between $z\bc(\k)$ and $z\bc(\k)+c$.

We first discuss the case of even $N$.  An example is sketched if
\fref{zflip}(b); there the symmetry is such that $\znom(\k)$ is
perfectly flat at $z\eee-c/2$, and the unit cell consists of the union
of the two shaded regions shown there.  Regarding the degeneracies
themselves, we let $\epsilon$ of \sref{shrink} go to zero faster than
$\delta\rightarrow0$, so that the cutting surface avoids the DRs. Then
$\tzo''$ reduces to the first term of \eq{tht}.  This in turn can be
broken into two contributions: one from the blue region
$z_n(\k)\in[\znom(\k)+\delta,\znom(\k)+c-\delta]$, and another from
the red region $z_n(\k)\in[\znom(\k)+c-\delta,\znom(\k)+c+\delta]$, in
\fref{zflip}(b).  The first region is centered about $z\eee0$, so all
contributions cancel in pairs, while the latter region vanishes in the
limit $\delta\rightarrow0$, yielding a vanishing contribution
  to $\tzo''$.\footnote{In the immediate vicinity of a Dirac node
  between a pair of Wannier bands, the Berry curvature arising from the
  coupling between the two crossing bands vanishes (although it
  diverges exactly at the node). Hence, the integrand in the first
  term of \eq{tht} is finite in the red region of
  \fref{zflip}(b) (excluding the DRs around the Dirac nodes),
  justifying the claim that the integral vanishes in the limit
  $\delta\rightarrow 0$.}
Thus, $\tzo''\eee0$.
We argued earlier that the $\tdxy$-type terms always
vanish in the presence of axion-odd symmetries,
so in fact the first two terms in \eq{th-d} both vanish, and we are
left with
\beq
\theta= \lim_{\delta\rightarrow0} \; \phi\bc \qquad\hbox{(even $N$)}
\eqlab{evencase}
\eeq
That is, the axion coupling is given by the sum of Berry phases of
the vanishingly small loops of intersection of the Wannier bands
with the cutting surface of \eq{shift} around the nodal points
as $\delta\rightarrow0$.  If these are
simple Dirac nodes lying on $\znom(\k)$, then we
know that each such Berry phase is $\pi$, and it follows that
the system is axion-odd if and only if the number of such
Dirac nodes is odd, as in \fref{zflip}(b).

If the number $N$ of degenerate bands at $(-c/2,\bar{\Gamma})$ is odd,
then arguments like those above yield
\beq
\theta= \lim_{\delta\rightarrow0} \; \phi\bc
   -\pi C_j \qquad\hbox{(odd $N$)}\,,
\eqlab{oddcase}
\eeq
where $C_j$ is the Chern number of the central band.
Again, if the nodes are simple Dirac nodes with their nodal points
lying on $z_j(\k)$, the system is
axion-odd if and only if the sum of this central Chern index and the
number of Dirac cones is odd.
Note that we could equally well have chosen to work with a nominal
cell boundary centered around $z\eee0$ instead of $z\eee-c/2$;
as in the disconnected case, this follows from the freedom to
shift the origin by $c/2$ along $z$.

\subsubsection{Symmorphic operation preserving $\zhat$}
\seclabel{simple}

Here we consider the case that $g$ preserves the sense of $\zhat$, so
$\gp=g$ has to be either TR itself, a time-reversed rotation about the
$z$ axis, or a simple reflection about a plane containing this axis,
as summarized in Table~\ref{tab:g}.  For the moment we assume there is
no associated fractional translation along $\zhat$;
that case will be considered in the next subsection.
Now the Berry curvature contribution from an area
element at $z$ on one Wannier band gets mapped onto one of opposite sign
on the same sheet at the same~$z$, giving canceling contributions to
$\tzo$.

In the disconnected case, this implies that $\theta\eee0$ and the
system is always in the axion-even phase, since any gap can be chosen
as the primary one defining the unit cell.  Conversely, if the system
is axion-odd, then the Wannier bands must be
connected.

For the connected case, any cutting surface $z\bc(\k)$
that respects the in-plane symmetries will automatically give
$\tzo''=0$,
since the same kind of cancellations occur.  In this case, the axion
$\zt$ index is just given by $\phi\bc$ in \eq{th-d}.
We can arbitrarily choose one
Wannier band $j$, make the cut at $z_{j}(\k)+\delta$ for vanishingly small
$\delta$, and count the total Berry phase of the small loops of
intersection of the Wannier bands with $z\bc(\k)$. In the case that these
are simple Dirac nodes, the system is axion-odd if and only if their
number is odd.

\subsubsection{Nonsymmorphic operation preserving $\zhat$}
\seclabel{ns}

Finally we again consider the $\zhat$-preserving case, but now
assuming that the corresponding space-group operation involves a
fractional translation $c/m$ along $\zhat$ for some integer $m$.
These can be glide mirrors, time-reversed half translations, and
time-reversed screws, denoted as $\ns{M_d}{2}$, $\ns{E'}{2}$, and
$\ns{C'_n}{m}$, respectively.\footnote{While $\ns{C_3}{2}$ and
  $\{C_3|\pm c/6\}$ are disallowed as elements of a nonmagnetic space
  group, since their cubes correspond to pure half-translations, the
  corresponding TR-composed operations are allowed in magnetic space
  groups.}

\paragraph{Fractional translations of $c/2$.}
%
For the moment, we focus on the operations that involve a
half-lattice-vector translation, i.e.,
$\ns{M_d}{2}$,
$\ns{E'}{2}$,
$\ns{C'_2}{2}$,
$\ns{C'_3}{2}$,
$\ns{C'_4}{2}$,
and $\ns{C'_6}{2}$.
Then, according to \eq{z-transf}, a Wannier band at $(z,\k)$ always has a
partner at $(z+c/2,\pm g_\perp\k)$, with the minus sign applying when
$g_\perp$ involves TR.  Moreover, all of the operations $M_d$, $E'$,
$C'_2$, $C'_3$, $C'_4$, and $C'_6$ reverse the sign of the
Berry curvature $\Ot$, which is thus equal in magnitude but opposite
in sign between these two partners.

If the band structure is disconnected, then there must be an even
number of gaps.  We choose a primary gap and divide the Wannier bands into
two disconnected groups related to each other by \eq{z-transf}: a
``lower group'' consisting of the first $J/2$ bands above the primary
gap, and an ``upper group'' consisting of the rest.  Consider a
band~$n$ in the lower group; its contribution $z_n(\k)\Ot_{nn}(\k)$ to
the integral in \eq{th-cg-b} can be paired with a contribution
\beq
z_{n'}(\k')\Ot_{n'n'}(\k')
  = \big[z_{n}(\k)+\frac{c}{2}\big]\big[-\Ot_{nn}(\k)\big]
\eqlab{twog}
\eeq
at $\k'=\pm g_\perp\k$
in the upper group.  The sum of these two contributions,
when combined with the $-1/c$ prefactor of \eq{th-cg-b},
yields $\Ot_{nn}(\k)/2$.
Since we know that the $\tdxy'$ term does not contribute,
\eqr{th-b}{thta} then yield
\beq
\theta=\frac{1}{2}\Phi_{\rm LG} =\pi C_{\rm LG}\,,
\eqlab{CLG}
\eeq
where $\Phi_{\rm LG}$ is the total flux of Berry curvature in the lower
group, with $C_{\rm LG}$ being the corresponding Chern number (an integer).
The system is thus axion-odd if and only if this Chern index is odd.

To prepare for the connected case, let us return to the disconnected
case for a moment.  We start with a cutting surface
$z_{\rm cut}^{(0)}(\k)$ that
lies entirely in the primary gap, so that the lower group has the
total Berry flux $\Phi_{\rm LG}$ of \eq{CLG}, which we now relabel as
$\Phi^{(0)}_{\rm LG}$.  We then raise the cutting surface to some
chosen new location $z\bc(\k)$ that cuts through some of
the bands at the bottom of the lower group.  (We also insist
that $z\bc(\k)$ avoids any DRs.)
The new lower group now extends from $z\bc(\k)$ to $z\bc(\k)+c/2$.  As
a result, the total Berry flux $\Phi_{\rm LG}$
in this group changes for two reasons: because of the omission of
terms below $z\bc(\k)$, which is compensated by the addition of a term
$\phi\bc$; and by the addition of new contributions near
$z\bc(\k)+c/2$, which is compensated by the addition of the
\textit{same} $\phi\bc$ because of the sign reversal of $\Ot_{nn}$.
Thus, we find that
\beq
\Phi^{(0)}_{\rm LG} = \Phi_{\rm LG}+2\phi\bc \,.
\eeq
The quantity on the right side of this equation is $2\pi$ times
an integer, even though the individual terms are not, and
the system is axion-odd if and only if this integer is odd.

Finally we argue that this same formula applies to the connected
case, since following in the spirit of the discussion in
\sref{cutsurf}, we can imagine that we introduce a perturbation
that opens a gap, place the cutting surface there, raise the cutting
surface, and then close the gap. A concise representation of the final
result is
\beq
\theta=\frac{1}{2}\Phi_{\rm LG}+\phi\bc \,.
\eqlab{tLG}
\eeq
This formula should be correct for any choice of cutting surface,
as long as it avoids the DRs.

In the case of Dirac cones connecting a pair of bands, we can choose a
cutting surface just above the average of these bands, so that
$\phi\bc$ could again be obtained as $\pi$ times the number of such
Dirac cones.  However, unlike in Sections~\ref{sec:rev} and
\ref{sec:simple}, here we cannot avoid doing an explicit calculation
of the total Berry flux in half the unit cell.  In this sense,
\eq{tLG} is not quite as convenient as our earlier connected-case
formulas.

\paragraph{Smaller fractional translations.}
%
We now consider the operations involving smaller fractional
translations, namely
$\ns{C'_3}{3}$,
$\ns{C'_6}{3}$,
$\ns{C'_4}{4}$,
$\ns{C'_3}{6}$, and
$\ns{C'_6}{6}$.
(We omit left-handed screws, since they behave in the same way as
right-handed ones.)

We start by assuming a disconnected band structure, and
consider the example of $\ns{C'_4}{4}$.  The number of gaps
must be a multiple of four; we choose one such gap to define
the unit cell in the usual way.  Now $\ns{C_2}{2}$,
which is the square of $\ns{C'_4}{4}$, is also in the symmetry group.
It simply translates the Wannier bands by a half lattice vector along
$\zhat$ with an extra $C_2$ rotation without affecting the value
of the Berry curvature on a given patch of the Wannier band, which is
just carried along to its new location.  From the point of view of
\eqo{tzo}{th-cg-b} for the contribution to $\theta$, the
rotation component is irrelevant, since it does not affect the
$z_n$ or $\Omega_{nn}$ value.
Thus, for our purposes we can think of $\ns{C_2}{2}$ as defining
a smaller ``unit subcell'' of height $c'=c/2$, and we can
compute $\theta$ by focusing on just one subcell containing
$J'=J/2$ Wannier bands.

For the case of isolated bands, for example, \eq{tzo} becomes
\beq
\tzo=-\frac{1}{c'}\int d^2k \sum_{n=1}^{J'} z_n\O_{nn}
\eqlab{tzo-half}
\eeq
(the fact that we count half as many bands is compensated by
the factor of two from the replacement of $c$ by $c'$ in the
prefactor), and a similar modification would apply to
\eqs{th-cg}{thta} in the case of composite groups.
Now the action of $\ns{C'_4}{4}$ in the subcell is entirely
analogous to the action of $\ns{C'_2}{2}$ in the full cell;
this is one of the cases that we studied above, and the same
conclusions apply.  That is, we conclude that $\theta$ is
given by \eq{CLG}, where $C_{\rm LG}$ is now interpreted
as the total Chern number coming from the bottom half of
the subcell, i.e., from the first $J/4$ bands.  The case
of a connected band structure under $\ns{C'_4}{4}$ is
handled using the same analogy to the $\ns{C'_2}{2}$ case,
and a formula like \eq{tLG} applies.

The same strategy can be applied to reduce the remaining
operations to previously studied ones in a similar way.
All of these remaining operations, namely
$\ns{C'_3}{3}$,
$\ns{C'_6}{3}$,
$\ns{C'_3}{6}$, and
$\ns{C'_6}{6}$,
have the property that their fourth power is (modulo full
translations) just a simple 3-fold
screw, which divides the unit cell into three subcells.
Thus, we can restrict our attention to a subcell of height
$c'=c/3$ containing $J'=J/3$ bands.

For the first two operations, $\ns{C'_3}{3}$ and $\ns{C'_6}{3}$,
there are no remaining fractional translations within the subcell.
Their third powers correspond to $E'$ and $C'_2$ respectively, so in
these cases the analysis of the subcell is analogous to the
symmorphic case discussed in \sref{simple}, and
the conclusions found there apply. Specifically, $\theta$ is trivial
in the disconnected case, and a counting of Dirac nodes can provide
the value of $\theta$ in the connected case.

For the last two operations, $\ns{C'_3}{6}$ and $\ns{C'_6}{6}$,
the subcell of height $c/3$ is subdivided by a further half
translation of $c/6$. The situation is similar to the $\ns{C'_4}{4}$
case discussed above, except that now the subcell is of size $c/3$
instead of $c/2$.  This subcell has lower and upper portions
containing $J/6$ Wannier bands each, related to each other by an
operation containing TR.  Thus, the analysis of the subcell in these
cases is analogous to the that of the full cell in the nonsymmorphic
$\ns{E'}{2}$ and $\ns{C'_2}{2}$ cases respectively. In the
disconnected case, \eq{CLG} applies with $C_{\rm LG}$ interpreted as
the Chern number of the first $J/6$ bands, and in the connected case
a formula like \eq{tLG} once again applies.%
\footnote{Alternatively, \eq{CLG} or \eq{tLG} can be applied directly
  in the full cell using the previous results for
  $\ns{E'}{2}$ or $\ns{C'_2}{2}$, the third powers of
  $\ns{C'_3}{6}$ and $\ns{C'_6}{6}$ respectively.}

\subsubsection{Summary}

In summary, for the case of a disconnected band structure, the rules
for determining the axion index are quite simple.  For a
$z$-reversing symmetry operator, we test for the presence of a
boundary-centered group with an odd Chern number, or equivalently, an
origin-centered one; if present, the system is axion-odd.  For
symmetry operators that preserve the sign of $\zhat$, we can consider
three cases.  If there is no associated fractional translation along
$\zhat$, the system is forced to be axion-trivial.  For the case of a
half translation, the system is axion-odd if and only if the total
Chern number in a half unit cell is odd.  For smaller fractional
translations, we can define a subcell that tiles the full cell under
simple screw rotations, and apply the earlier analysis to the subcell.

The rules for the connected case are more complicated, but in the
simplest case that the connection is via Dirac nodes, the results
can be expressed in terms of a counting of Dirac nodes,
and may also require the calculation of the total Berry flux
of a band or subgroup of Wannier bands in the unit cell.

Note that we may have some choice regarding the crystallographic
direction along which to perform the wannierization.  In the case of
$I$ or $E'$, any convenient primitive reciprocal direction will do.
In the case of a mirror, we may choose either an axis normal to or
lying in the mirror plane, applying the analysis of
Secs.~\ref{sec:rev} or \ref{sec:simple} respectively.  (If
$\theta\!=\!\pi$, the Wannier bands are necessarily connected for the
choice of wannierization in the mirror plane.)  The case of $C'_2$
rotations provides a similar choice of options, i.e., the HW
construction can be done either normal to the $C_2$ axis
(Sec.~\ref{sec:rev}) or along it (Sec.~\ref{sec:simple}).  Of course,
there may be more than one axion-odd symmetry in the magnetic point
group, in which case one also has a freedom to choose which operation
to select.  Thus, it may often be possible to simplify the
determination of the axion index by an appropriate choice of symmetry
and setting.

\subsection{Symmetry conditions for topologically-protected spectral flow}
\seclabel{flow}

Here, we address two closely related questions concerning
spectral flow in an axion-odd insulator.  First, we identify the
symmetry conditions under which topologically protected metallic
states must exist on a given surface.  Conversely,
when are insulating surfaces allowed to appear?
We then discuss the conditions under which the flow of the
bulk Wannier bands in a given direction is topologically
protected.  Conversely, when can
an axion-odd insulator have a gapped Wannier band structure?

\subsubsection{Protected flow of the surface energy bands}
\seclabel{flow-surf}

To answer the first question posed above, we ask whether the surface
magnetic point group contains any axion-odd symmetry operations.  If
so, the surface AHC would be forced to vanish (see below).  However,
this is inconsistent with the half-integer surface AHC that,
according to \eq{surfahc}, must occur for any insulating
surface of our
assumed axion-odd bulk insulator.  Thus, under these conditions
the surface is necessarily metallic.  In this case,
\eq{surfahc} becomes\cite{vanderbilt-book18}
\beq
\sigma^{\text{surf}}_{\text{AHC}} =
\frac{e^2}{h}\frac{\phi-\theta}{2\pi} \ \text{mod} \ e^2/h \, ,
\eqlab{surfahc-metallic}
\eeq
where the extra phase angle $\phi$ is the Berry phase summed over
all the Fermi loops in the surface BZ. Since each Dirac cone
contributes a Berry phase of $\pi$, a topological metallic surface
must have an odd number of them to cancel the $\theta=\pi$
contribution from the bulk.

If no axion-odd symmetries are present at the surface, then
the surface is allowed to be insulating.  Of
course, whether it is really insulating or not will depend on
details of the surface electronic structure for the particular
surface termination, but the bulk topology no longer requires a
metallic state.

One immediate consequence is that if TR itself is a symmetry of the
bulk, and also of the surface of interest, then this surface must be
metallic.  This is the well-known statement that strong TIs have
metallic surface states on all surfaces, unless TR is somehow
broken on the surface.  By contrast, if inversion is the only
axion-odd symmetry protecting the axionic topology of the bulk, then
all surfaces are allowed to be insulating, because no surface
preserves inversion symmetry.

More complicated situations can be analyzed as follows.  For a given
facet on the surface of an axion-odd crystal, we first reduce from the
3D bulk magnetic space group to the 2D magnetic space group that
describes the highest symmetry that the surface could possibly have.
This will consist of all operations of the 3D group that leave the
surface normal (chosen as $\zhat$) invariant, and that involve no
translation (either by a full or a fractional lattice vector) along
$\zhat$; the axion-odd operations among these, if any, are of the
$z$-preserving type.  We then construct the surface magnetic point
group in the usual way, by listing the point operations appearing in
the space group.  Finally, we omit any operations that are not
axion-odd. The remaining operations that may be present are the
$z$-preserving operations in Table~\ref{tab:g}: TR itself, the $C'_n$
time-reversed rotations about $\zhat$ ($n\,=\,\{2,3,4,6\}$), and
reflections $M_d$ about a plane whose unit normal lies in the surface.
All these symmetries flip the sign of the surface AHC (a pseudovector
pointing along $\zhat$), forcing it to vanish. Hence, if any of these
symmetries are present at the surface, then the surface must be
metallic, with $\phi=\pi$ in \eq{surfahc-metallic}.

Of course, any given surface may have lower symmetry than the
one allowed by the above considerations.  For example, a bulk-allowed
$M_d$ mirror symmetry may be spontaneously broken by the
formation of a symmetry-lowering surface reconstruction, or TR
may be broken at the surface by the spontaneous appearance of
magnetic order. In general, if any of the
axion-odd symmetries survive at the surface,
it must be metallic; otherwise it can be insulating.

We mentioned earlier that if inversion is the only axion-odd bulk
symmetry, then all surfaces are allowed to be insulating.  One may ask
whether this is also true for any other symmetries.  The answer is
yes. Consider, for example, the case that an improper
rotation $S_3$, $S_4$, or $S_6$ about the $z$ axis is a
symmetry; then for any facet normal $\nhat$, the
 component of $\nhat$ on the $(x,y)$
plane is rotated by the symmetry, and the $z$ component is
reversed, so there are no surfaces that obey this symmetry.  The
time-reversed screw rotations also have this property. This time they
do preserve $\nhat$ for surfaces normal to the rotation axis, but the
fractional translation that comes with the screw operation is always
inconsistent with such a surface.

\subsubsection{Protected flow of the bulk Wannier bands}
\seclabel{flow-wannier}

As we have seen, a surface of unit normal $\zhat$
on an axion-odd insulator is required to be metallic
if any bulk axion-odd symmetries of the $z$-preserving
type are preserved at the surface.  Assume that such a
high-symmetry metallic surface has been prepared.  Following
Refs.~[\onlinecite{fidkowski-prl11,neupert-bookchapter18}], the
surface spectrum can be continuously deformed into the bulk Wannier
spectrum obtained by wannierizing along $\zhat$.  Briefly, the actual
surface band structure is connected to a model surface obtained by
energetic flattening and spatial truncation, and then the abrupt
truncation is replaced by a crossover region whose width is allowed to
diverge.  In this limit, these authors show that the Wannier band
structure is recovered.  Since the surface spectrum flows by
hypothesis, the Wannier spectrum must flow as well.

This argument, based on the bulk-boundary correspondence, reproduces
the conclusions of \sref{simple}, where purely bulk considerations
led to the conclusion that the Wannier bands must be
connected if a $z$-preserving axion-odd symmetry is present.
These considerations also imply that the
flow should occur via an odd number of Dirac nodes between adjacent
Wannier bands.

We emphasize, however, that the bulk-boundary argument does not
necessarily work in reverse.  That is, if the surface band
structure does not flow, this does not necessarily imply the same
for the bulk Wannier bands.  Sometimes a $z$-reversing
or $z$-nonsymmorphic symmetry can lead to flow of the Wannier
bands in a ``fragile'' sense.  By this we mean that the flow can be
destroyed, without breaking any axion-odd symmetries, by adding
some weakly-coupled trivial bands to the valence manifold; this
behavior was demonstrated in Ref.~[\onlinecite{wieder-arxiv18}]
for the case of inversion symmetry.  This is again consistent
with the analysis of \srefs{rev}{ns}, where both connected and
disconnected Wannier band structures were found to be compatible
with $\theta=\pi$.
This will be illustrated for the case of a pyrochlore model with
inversion symmetry in \sref{pyro}, and for the case of a
model with glide mirror symmetry in \sref{glidemodel}. Since
$z$-reversing and $z$-nonsymmorphic operations are never symmetries of
the surface, there is never any such fragile protection of the flow of
the surface energy bands in these cases.

\section{Special cases and numerical tests}
\seclabel{cases}

In this section we consider four axion-odd symmetry operations that
deserve special emphasis: TR (i.e., $E'$); inversion $I$; mirror (either $M_z$ or $M_d$, depending on the wannierization direction);
and glide mirror $\{M_d|c/2\}$. In each case we shall
introduce an elementary tight-binding model that can be used to
illustrate some of the behaviors expected from theory.

\subsection{Time-reversal symmetry}
\seclabel{TR}

\subsubsection{General considerations}

Here we consider the case that TR itself, $E'$, is a symmetry of the
system.  This is a $z$-preserving symmetry as discussed in
\sref{simple}, and the conclusions given there apply directly.

We first consider the case of spinor electrons, so that
$\left(E'\right)^2=-1$.  As is well known, all energy bands of a 2D
system appear as Kramers pairs at all four TRIM.  In a similar way,
the Wannier bands of a 3D system all appear as Kramers degeneracies at the
four PTRIM, as discussed in
Ref.~[\onlinecite{taherinejad-prb14}].  Therefore, any given Wannier band
$j$ is either Kramers degenerate with band $j-1$, or with band $j+1$,
at each of the four PTRIM.  Let us call these ``down-touchings'' and
``up-touchings'' respectively; their numbers are $\Nd(j)$ and
$\Nu(j) =4-\Nd(j)$ respectively. The next band clearly has
$\Nd(j+1)= \Nu(j)$, etc.  Since $\Nd(j+1)=4-\Nd(j)$, there are three
cases to consider:

\begin{itemize}

\item $\Nd$ alternates between 0 and 4 as bands are counted.
The Wannier bands are thus ``glued together'' in pairs, and are generically
gapped from the next higher or next lower pair.\footnote{As an
  exception, if other symmetries are also present,
  these could induce additional degeneracies between Wannier bands,
  typically at high-symmetry points other than the TRIM.}
From this it follows that the axion $\zt$ index is trivial.  As
discussed in Ref.~[\onlinecite{taherinejad-prb14}], it also follows
that the $\zt$ indices are trivial on the four TR-invariant planes
$k_x=0$, $k_x=\pi$, $k_y=0$, and $k_y=\pi$ in the 3DBZ.
This system is either topologically trivial, or it is a weak
topological insulator with indices (0;001), i.e., equivalent to a
stack of quantum spin Hall (QSH) layers along $z$.

\item $\Nd=2$ for all bands.  The Wannier band structure is connected;
there are no gaps.  These systems correspond to weak
topological insulator configurations, this time corresponding to
a stacking of QSH layers along $x$ or $y$ or in the diagonal $xy$ direction.
According to the discussion in \sref{simple}, we can choose
a cutting surface just above any chosen band;
the two Dirac cones cut by this surface contribute an even integer
times $\pi$, and the system is again axion-trivial.

\item $\Nd$ alternates between 1 and 3 as bands are counted.
The Wannier band structure is again connected.
A cutting surface is again chosen just above one Wannier band;
since there will
be 1 or 3 Dirac cones cut by this surface, each carrying a Berry
phase of $\pi$, the system is a strong topological insulator and
it is axion-odd.

\end{itemize}
These conclusions are consistent with the well-known properties
of TR-protected topological insulators, as discussed, for example,
in Ref.~[\onlinecite{taherinejad-prb14}].

For the case of scalar particles, $\left(E'\right)^2=+1$,
the Kramers degeneracies are not enforced, and the Wannier band structure
will generically be disconnected, in which case $\theta\eee0$.  In
case a connected Wannier band structure is enforced by degeneracies
associated with additional symmetries, the Dirac node counting
procedure described at the end of \sref{simple} should still apply.

\subsubsection{Fu-Kane-Mele model}
\seclabel{fkm-stuff}

To illustrate how the trivial, weak-TI, and strong-TI phases manifest
themselves in the HW representation, we use the Fu-Kane-Mele (FKM)
model.~\cite{fu-prl07} In this model, spinor $s$-like orbitals
interact with spin-dependent and spin-independent hoppings on a
diamond lattice of conventional lattice constant $a$. The Hamiltonian
contains two terms,
\beq
H = \sum_{\langle ij \rangle}  t_{ij}c^\dagger_{i}c_{j} +
\lambda_{\rm SO}\sum_{ \la ij \ra} ic^\dagger_{i} \,
\boldsymbol{g}_{ij} \cdot \boldsymbol{\sigma} \, c_{j}\,.
\eqlab{FKM}
\eeq
The first term describes spin-independent first-neighbor hoppings,
denoted as $\langle ij \rangle$; the hopping amplitudes are
$t_{ij} = t_0$ for all bonds except those aligned along $[111]$, for
which $t_{ij}=t_0 +\Delta t$. The second term, which models the effect
of spin-orbit coupling, consists of spin-dependent second-neighbor
hoppings denoted as $\la ij \ra$;
$\boldsymbol{\sigma}=(\sigma_x,\sigma_y,\sigma_z)$ are the Pauli
matrices, and
$\boldsymbol{g}_{ij} = \boldsymbol{d}_{il} \times \boldsymbol{d}_{lj}$
where $\boldsymbol{d}_{il}$ and $\boldsymbol{d}_{lj}$ are
the bond vectors of the first-neighbor pairs making up the
second-neighbor bond.
We set $t_0=1$ and $\lambda_{\textrm{SO}} = 0.125$, and adjust $\Delta t$ according to the phase diagram
\[
\Delta t \in
    \begin{cases}
        (-\infty,-4),& \text{Trivial insulator}\\
        (-4,-2),     & \text{Strong TI} \\
        (-2, 0),     & \text{Weak TI} \\
        (0, 2),     & \text{Strong TI} \\
        (2,\infty),     & \text{Trivial insulator} 
    \end{cases}
\]
to put the system in a trivial-insulator, weak-TI, or strong-TI
phase at half filling.\cite{taherinejad-prb14}

\begin{figure}
\centering\includegraphics[width=\columnwidth]{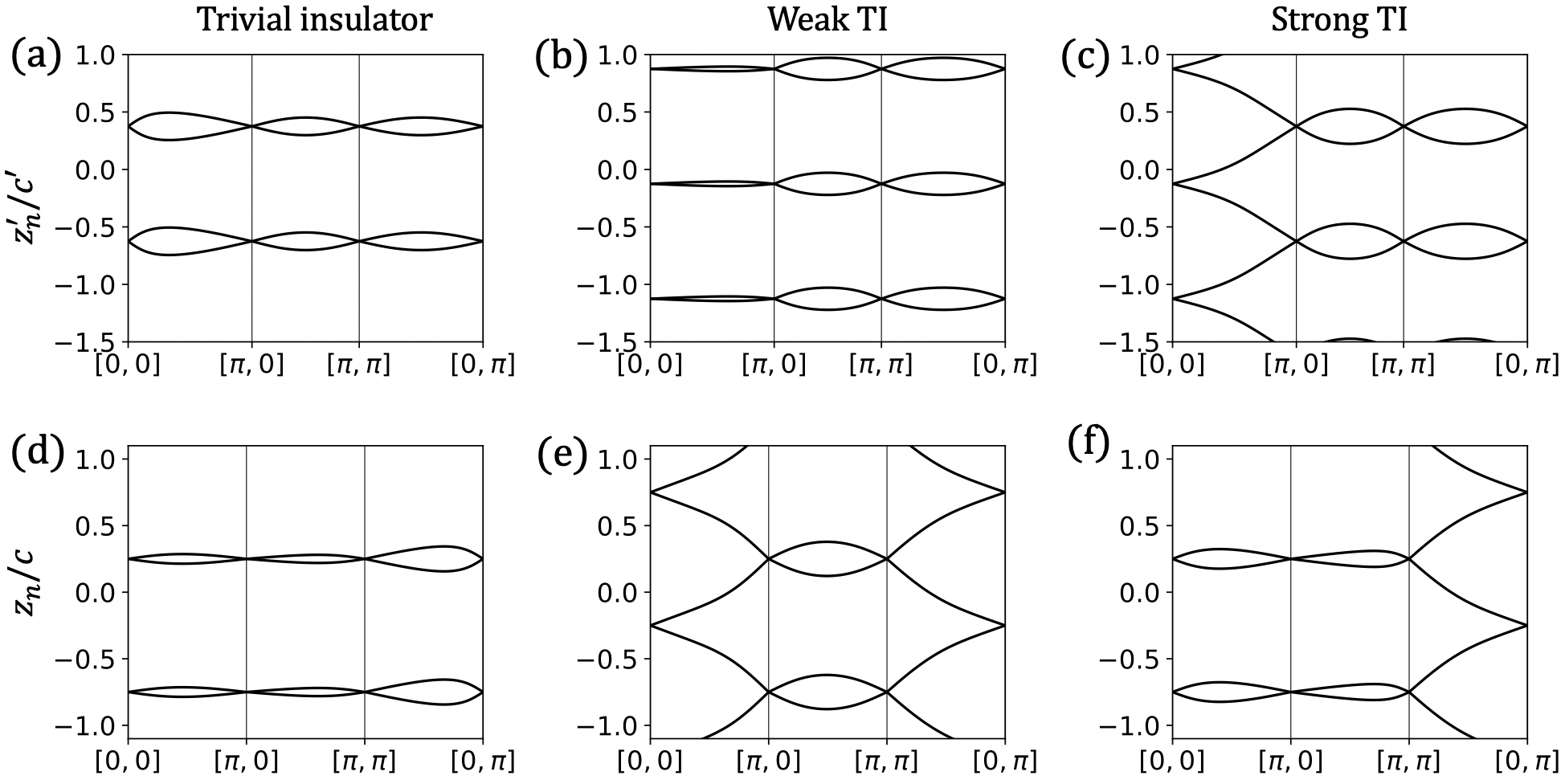}
\caption{Wannier band structures for the FKM model in a
  trivial-insulator phase with $\Delta t=2.5$ (a,d), a weak-TI phase
  with $\Delta t=-1.0$ (b,e), and a strong-TI phase with
  $\Delta t=1.0$ (c,f). In (a-c) the wannierization is along the (111)
  direction, denoted as $\zhat'$, with lattice periodicity
  $c'=a\sqrt{3}/3$; in (d-f) we wannierize along (001), with lattice
  periodicity $c=a/2$.  The bands are plotted along high-symmetry
  lines connecting the four PTRIM in the 2DBZ. }
\label{fig:fkm}
\end{figure}

For our numerical tests we pick a trivial insulator with
$\Delta=2.5$, a weak TI with $\Delta t=-1.0$, and a strong TI with
$\Delta t=1.0$.  Their Wannier band structures are shown in \fref{fkm};
in the top panels the wannierization is along (111), and in the bottom 
panels it is along (001).

As explained earlier, in a trivial insulator the number $\Nd$ of
Kramers down-touchings at the PTRIM is expected to alternate between 0
and 4 irrespective of the wannierization direction, and this is indeed
what is observed in the left panels of \fref{fkm}.

As for the middle panels of \fref{fkm}, we note that a weak TI can
always be thought of as a stack of QSH layers in some direction
determined by the weak indices.\cite{fu-prl07} In this example the
weak-TI phase corresponds to a stack of QSH layers along (111), which
should not come as a surprise since we have weakened the hoppings
along [111]. Hence, when wannierizing along (111) we expect a gapped
spectrum with $\Nd$ alternating between 0 and 4, as seen in
\fref{fkm}(b). On the other hand, for other wannierization directions
we expect a connected spectrum with $\Nd=2$ for all bands, as seen in
\fref{fkm}(e).

Finally, the behavior of the right panels of \fref{fkm}
is as predicted for a strong TI: the Wannier band structure is necessarily
connected, with $\Nd$ alternating between one and three.

\subsection{Inversion symmetry}
\seclabel{I}

\subsubsection{General considerations}

Next we consider the case that a simple inversion $I$ is a symmetry of
the system, so that it is denoted as an ``axion insulator'' if it is
axion-odd.  This is an example of a $z$-reversing axion-odd symmetry,
and the conclusions derived in \sref{rev} apply. To summarize, we
found that in the disconnected case, $\theta=\pi$ if and only if there
is a boundary-centered group and its Chern number $C\bbc$ is odd. (The
same applies to the origin-centered group.)  Unlike TR, the presence
of inversion symmetry does not require an axion-odd insulator to have
a connected Wannier band structure,\cite{varnava-prb18,wieder-arxiv18}
as discussed in \sref{flow-wannier}.  Regardless of whether it does or
not, we can frequently deduce the $\zt$ index from a node-counting
argument, e.g., by focusing on the set of Wannier bands that touch
$z\eee-c/2$ at $\bar{\Gamma}$.  If the number of bands in this group
is even, we place a cutting surface infinitesimally above the average
of the central two bands, and conclude that the $\zt$ index is odd if
and only if the number of Dirac nodes sliced in this way is odd.  In
case the number of Wannier bands is odd, then we need information
about the Chern number of the central band in addition.  The same
considerations apply to an analysis centered around $z\eee0$.
We note in passing that the work of Alexandradinata
et~al.\cite{alexandradinata-prb14} provides a complementary
view of the Wannier (Wilson-loop) band structure for 2D insulators.

The axion $\zt$ index can also be obtained from a very different
approach based on parity counting.  In a path-breaking work,
Turner {\it et~al}.\cite{turner-prb12} derived a set of rules for
deducing many topological properties of a centrosymmetric crystalline
material based on a counting of the odd-parity eigenvalues of the
occupied Bloch states at the eight TRIM in the 3DBZ.  They showed that
the system can be insulating only if the total number of odd-parity
states is an even integer, and moreover it is axion-even or axion-odd
depending whether this number is of the form $4n$ or $4n+2$, where $n$
is an integer.

In the \aref{parity-count}, we carry this analysis over to the
Wannier band structure.  Here, the original eight TRIM of the
3DBZ
project onto four PTRIM in the 2DBZ.  At each of these $\k$, the
Wannier bands can be classified as being of even or odd parity at $z\eee0$;
even or odd parity at $z\eee c/2$; or appearing in pairs at $\pm z$.
We develop counting rules inherited from the 3D parity counting of the
Bloch states. Consistent with the analysis above, we find that the
axion $\zt$ index can often be determined directly from an inspection
of the Wannier band structure, even without evaluating the parities of the
HW states.  In some cases, however, some parity eigenvalues do need to
be evaluated, allowing a determination of whether
the boundary-centered group has an odd Chern number as discussed above.
The details of this analysis are deferred to the \aref{parity-count}.

\subsubsection{Pyrochlore model}
\seclabel{pyro}

For the case of inversion symmetry we borrow an illustrative class of
centrosymmetric tight-binding models discussed in
Ref.~[\onlinecite{varnava-prb18}].
These models consist of
one pair of spinor basis orbitals
per site of the pyrochlore lattice, as might be used to describe the
$J_{\textrm{eff}} = 1/2$ manifold of the pyrochlore iridates. The
Hamiltonian reads
\beq
H = t \sum_{\langle ij \rangle}  c^\dagger_{i}c_{j} + 
\lambda \sum_{ \langle ij \rangle} i \sqrt{2} c^\dagger_{i} \, \boldsymbol{\hat{g}}_{ij}\cdot \boldsymbol{\sigma}\, c_{j} +
\Delta \sum_{i} \boldsymbol{\hat{n}}_i \cdot \boldsymbol{\sigma}c^\dagger_{i}c_{i} \;,
\eqlab{pyroH}
\eeq
where the spin indices are suppressed.  Just as for the FKM model, the
first two terms describe spin-independent and spin-dependent
first-neighbor hoppings, with amplitudes $t$
and $\lambda$ respectively.
In the latter, $\boldsymbol{\hat{g}}_{ij}$ is a unit vector normal to
both the hopping and the potential gradient directions; see
Ref.~[\onlinecite{varnava-prb18}] for details.  The third term is a
Zeeman term responsible for breaking TR
symmetry, with the vectors $ \boldsymbol{\hat{n}}_i$ ($i=1,2,3,4$)
describing the directions of the magnetic moments on the pyrochlore
lattice sites. We choose the all-in-all-out magnetic
configuration,\cite{varnava-prb18} and consider the model at half
filling.

Whether the system is axion-even or axion-odd can easily be determined
using the parity criteria for inversion-symmetric
insulators.\cite{turner-prb12} That is, the system is even or odd if
the total number of odd-parity Bloch states at the TRIM is of the form
$4n$ or $4n+2$ respectively, where $n$ is an integer.
Table~\ref{tab:parities} shows how the odd-parity states are
distributed among the eight TRIM for three representative
choices of parameters (see the caption) corresponding to different
regions of the phase diagram. Their total number is 12, 10, and 14,
indicating axion-even, axion-odd, and axion-odd topology,
respectively. We refer to them below as the trivial, Axion I, and Axion II
cases, respectively.

\begin{table}[t]
\begin{ruledtabular}
\caption{Number of odd-parity Bloch states at the eight TRIM for three
  representative cases: ``Trivial'' phase with
  $(t,\lambda,\Delta) = (1.0,0.1,3.0)$; ``Axion I'' phase with
  $(t,\lambda,\Delta) = (1.0,0.1,0.6)$; and ``Axion II'' phase with
  $(t,\lambda,\Delta) = (-1.0,-0.1,0.6)$.  The TRIM are labeled as in
  Ref.~[\onlinecite{wan-prb11}].}
\begin{tabular}{lccccc}
Phase & $\Gamma$ & $X,\,Y,\,Z$ & $L\,(\times 3)$&  $L'$ & Total  \\
\hline
Trivial            & 0& 2& 1& 3 & 12  \\
Axion I  & 0& 2& 0& 4 & 10  \\
Axion II & 0& 2& 2& 2 & 14  \\
\end{tabular}
\label{tab:parities}
\end{ruledtabular}
\end{table}

We present the corresponding Wannier bands in \fref{pyro}.  Along
$(111)$ the pyrochlore lattice consists of alternating triangular and
kagome layers containing one and three atoms, respectively.  This is
reflected in the Wannier band structure of the trivial phase in
\fref{pyro}(a), since in that phase electrons are localized on the
atomic sites.  Similarly, along $(001)$ we have a stack of tetragonal
layers rotated by $45^{\circ}$ with respect to each other, and in
\fref{pyro}(d) we see that the OC and BC groups have two Dirac nodes
each, implying $\theta\eee0$.

In the axion-insulator phase, the Wannier band structure can be connected or
disconnected depending on the choice of parameters, as illustrated in
the middle and right panels of \fref{pyro}.
The two axion phases differ by four odd-parity states (see
Table~\ref{tab:parities}), consistent with both being
axion-odd.  The fact that there is flow in the case of
Axion I state, but not for Axion II, is another illustration of the
kind of fragile protection that was discussed in \sref{flow-wannier}.
For example, it is evident  that the addition of a pair of trivial
flat Wannier bands near $z\eee\pm0.25c$ would convert the hybrid
Wannier band structure in \fref{pyro}(b) into a disconnected one
after hybridization with the crossing bands would occur.

In the connected Axion I case, the number of Dirac nodes at the nominal cell
boundary is an odd integer, as per \eq{evencase}. Indeed, in \fref{pyro}(b)
there are five nodes at $z=c/2$ and three at $z=0$, while in
\fref{pyro}(e) there are three nodes at both $z=c/2$ and $z=0$.

\begin{figure}
\centering\includegraphics[width=\columnwidth]{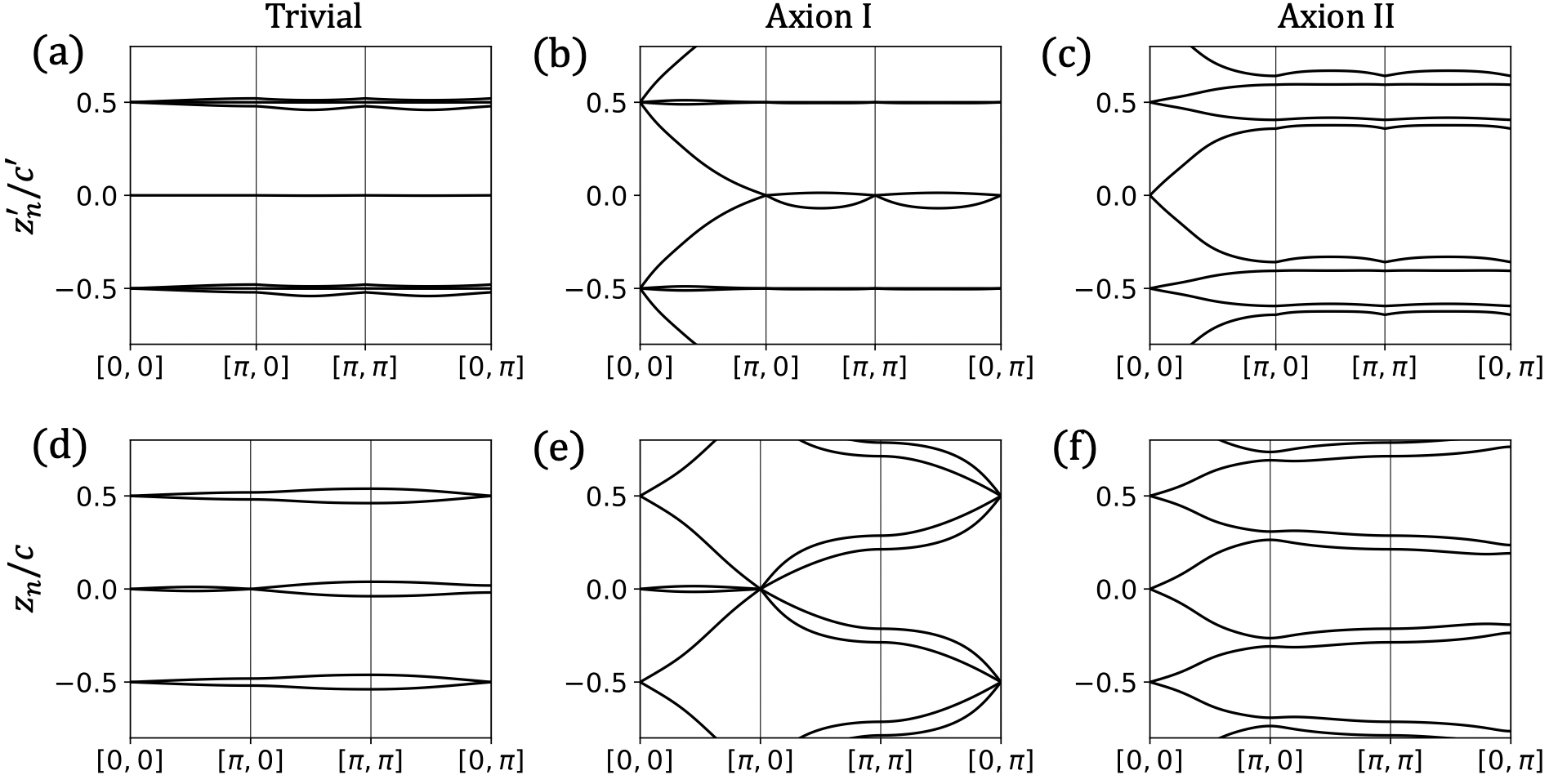}
\caption{Wannier band structures for the pyrochlore model, wannierized
  along (111) (denoted $\zhat'$) in (a-c), and along (001) in (d-f).
  (a,d) Trivial insulator; (b,e) Axion I phase; (c,f) Axion II phase
  (see caption of Table~\ref{tab:parities}).}
\label{fig:pyro}
\end{figure}

For the disconnected band structures in the left and right panels of
\fref{pyro}, we showed in \eq{half} that $\theta=0$ or~$\pi$ depending
on whether the Chern index of the boundary-centered group is even or
odd.  While it is possible in principle to compute Berry curvatures
and Chern numbers from the bulk Wannier
bands,\cite{taherinejad-prl15,olsen-prb17} here for convenience we
carry out the analysis based on a slab configuration chosen thick
enough that the central Wannier bands are bulk-like.  In particular, we
construct slabs with a thickness of ten unit cells along
$\zhat'\parallel (111)$.
Slab HW states maximally-localized along $\zhat'$ are obtained as
eigenstates of the projected position operator
$P_\k z'P_\k$.\cite{kivelson-prb82,marzari-prb97,olsen-prb17} For a
sufficiently thick slab, the ones located far from the surfaces are
indistinguishable from the bulk HW states.  We then organize the
corresponding eigenvalues --~the Wannier bands of the slab~--
into origin-centered and boundary-centered groups, i.e., centered at
integer and half-integer $z'/c'$, respectively. Finally, we calculate
explicitly the Chern index of each group of Wannier bands using
\eq{half}.  In \fref{cz_pyro} we confirm that in the disconnected Axion~II
phase the Chern index of the boundary-centered
groups is an odd integer ($C=-1$), while in the trivial phase it is
even ($C=0$).

For the Axion II phase, we could have reached the
same conclusion from either \fref{pyro}(c) or (f) by counting Dirac
nodes, since the origin centered and boundary centered groups have 
two Wannier bands each.
For the trivial phase this strategy cannot be used for the case
of 3+1 bands in \fref{pyro}(a), but it does does apply to the (001)
wannierization with 2+2 bands in \fref{pyro}(d), confirming the
trivial topology.

\begin{figure}[t]
\centering\includegraphics[width=\columnwidth]{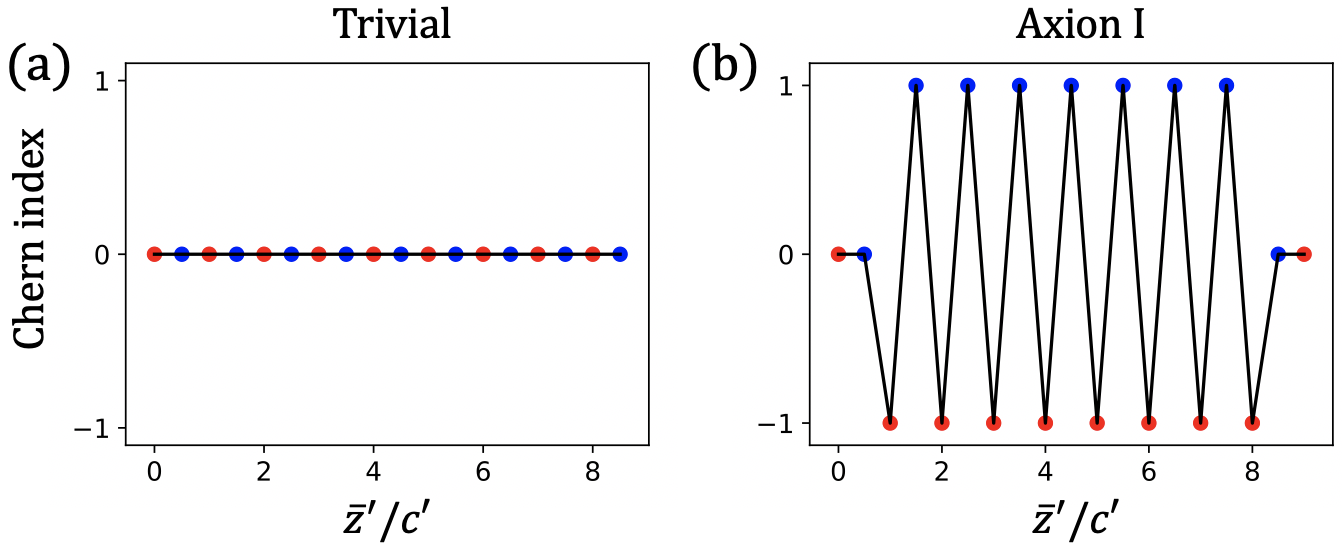}
\caption{Chern index for connected groups of slab Wannier bands
  of the pyrochlore model, plotted
  as a function of their average positions $\overline{z}'$.  Red and
  blue markers correspond to groups with integer and half-integer
  $\overline{z}' /c'$ respectively.  We consider slabs with a
  thickness of 10 unit cells along (111) in (a) the trivial phase of
  \fref{pyro}(a), and (b) the axion~II phase of
  \fref{pyro}(c); in the axion phase, the
  top layer was ``exfoliated'' to remove unwanted nontopological
  surface states.  In the trivial phase, both the groups with integer
  $\overline{z}'/c'$ --~one band~-- and those with half-integer
  $\overline{z}'/c'$ --~three bands~-- have zero Chern index. In the
  axion~II phase both groups comprise two bands each, and far from the
  surfaces their Chern indices are $-1$ and $+1$
  respectively.}
\label{fig:cz_pyro}
\end{figure}

\subsection{Mirror symmetry}
\seclabel{mirror}

\subsubsection{General considerations}

In \sref{gen-axion} we considered two
types of mirrors relative to a given wannierization direction
$\zhat$. When $\zhat$ lies on the mirror plane, the mirror is of
type $M_d$, and is a symmorphic $z$-preserving operation as discussed in \sref{simple}. When $\zhat$ is perpendicular to the mirror plane, we have
an $M_z$ mirror reflection, a $z$-reversing
axion-odd symmetry of the kind presented in \sref{rev}.

The case of $M_d$ is similar to that of TR, while the case of $M_z$
is similar to that of inversion, but with an important difference.
In the case of inversion, the Bloch states only have
well-defined parities at the eight TRIM, which are the
reciprocal-space points that are carried onto themselves by inversion.
By the same token, the parity analysis of the HW states discussed
above is only applicable at the four PTRIM.  As a result, Dirac
touchings between Wannier bands may occur at $z\eee0$ or $z\eee c/2$ at the
four PTRIM, but are generically absent elsewhere unless enforced by
additional symmetries.

The $M_z$ operation, on the other hand, acts as the identity on $k_x$
and $k_y$, so that all $\k$ in the 2DBZ follow ``mirror parity'' rules
similar to those discussed above for the PTRIM under inversion.  That
is, at any $\k$, the Wannier bands can be decomposed into those centered at
$z\eee0$ with even or odd parity labels; those centered at $z\eee c/2$
with even or odd parity labels about their center; and
mirror-symmetric pairs of Wannier bands at $\pm z$.  A consequence is that
Wannier bands can be pinned exactly at $z\eee0$ or $z\eee c/2$ over the
entire 2DBZ.

Moreover, in the presence of mirror symmetry, a new kind of
topological invariant known as the ``mirror Chern number'' comes into
play.\cite{teo-prb08,hsieh-nc12,ando-arccp15}
The mirror Chern number on the $k_z\eee0$ symmetry plane is normally
defined in reciprocal space in terms of the difference of Chern
indices of the even- and odd-parity mirror subspaces on this
plane, whose points are left invariant under $M_z$.  Except
for certain centered lattices,\cite{varjas-prb15} a second mirror
Chern number can be defined in the same way on the $k_z\eee \pi/c$
plane as well.

In a separate collaboration involving some of us, we have investigated
the Wannier bands in the presence of $M_z$ symmetry
in some detail, clarifying the generic behaviors that are expected,
and discussing the rules for deducing not only the axion $\zt$ index,
but also the mirror Chern number, from the Wannier band structure.  The
work will be published elsewhere.\cite{rauch-draft}

\subsubsection{Alternating Haldane model}

To illustrate the case of mirror symmetry we use a spinless
tight-binding model introduced in Ref.~[\onlinecite{olsen-prb17}],
which we refer to as the ``alternating Haldane model.''
This model consists of layers of the Haldane model\cite{haldane-prl88}
stacked along the $(001)$ direction, with two layers per cell. The Hamiltonian can be expressed as
\beq
H = \sum_p \Big[
H_p + t_{3,p}\sum_i \tau_i \big(c^\dagger_{p,i}c_{p+1,i} + \text{h.c.}\big)
\Big]
\eqlab{bihaldane}
\eeq
where the first term describes isolated layers, the second term
couples adjacent layers, and ``h.c.'' stands for ``Hermitian
conjugate.''  The Hamiltonian is constructed in such a way that in
the decoupled limit adjacent layers have
opposite Chern numbers. Specifically,
\beq
\begin{split}
H_{p} = (-1)^p \Delta \sum_i \tau_i c^\dagger_{p,i}c_{p,i} + 
t_1 \sum_{\langle ij \rangle} c^\dagger_{p,i}c_{p,j}  \\ 
+(-1)^p t_2 \sum_{\la ij \ra}i\nu_{ij}c^\dagger_{p,i}c_{p,j}
\eqlab{haldane}
\end{split}
\eeq
where indices $i$ and $j$ label the honeycomb sites and
$\tau_i = +1(-1)$ for $i \in A(B)$, where $A$ and $B$ are the two
honeycomb sublattices.  The first and second terms contain the on-site
energies and nearest-neighbor hoppings respectively, and the third
describes a pattern of staggered magnetic fluxes generated by complex
second-neighbor hoppings. Therein, $\nu_{ij}=+1$ ($-1$) if
the hopping direction from $j$ to $i$ is right-handed (left-handed)
around the center of a plaquette.\cite{haldane-prl88}
The $(-1)^p$ factor in the third term is
responsible for reversing the Chern number between adjacent
layers.  The first term also contains a $(-1)^p$ factor that
reverses the on-site energies; as a result, adjacent layers are
not simply time-reversed as in the case of ``antiferromagnetic
TIs'' with $\ns{E'}{2}$ symmetry.\cite{mong-prb10}

The above model was used in Ref.~[\onlinecite{olsen-prb17}] to pump a
quantum of axion coupling during a slow cyclic evolution of the
Hamiltonian along the path
\beq
\begin{split}
    t_1 & = 4.0\,, \\
    t_2 & = -1.0\,, \\
    t_{3,p} & = -(1+(-1)^{p-1}0.4\sin\phi)\,, \\
    \Delta & = -(3\sqrt{3}+2\cos\phi)\,,
\end{split}
\eqlab{phi}
\eeq
parameterized by the angle $\phi$.  Here we are interested in the
configurations $\phi=0,\pi$ where $t_{3,p}$ becomes independent of
$p$, and as a result the model acquires mirror
symmetry about the planes of the layers.  At $\phi\eee 0$ the
half-filled system is axion-even, and at $\phi\eee\pi$ it is
axion-odd.\cite{olsen-prb17}

\begin{figure}
\centering\includegraphics[width=\columnwidth]{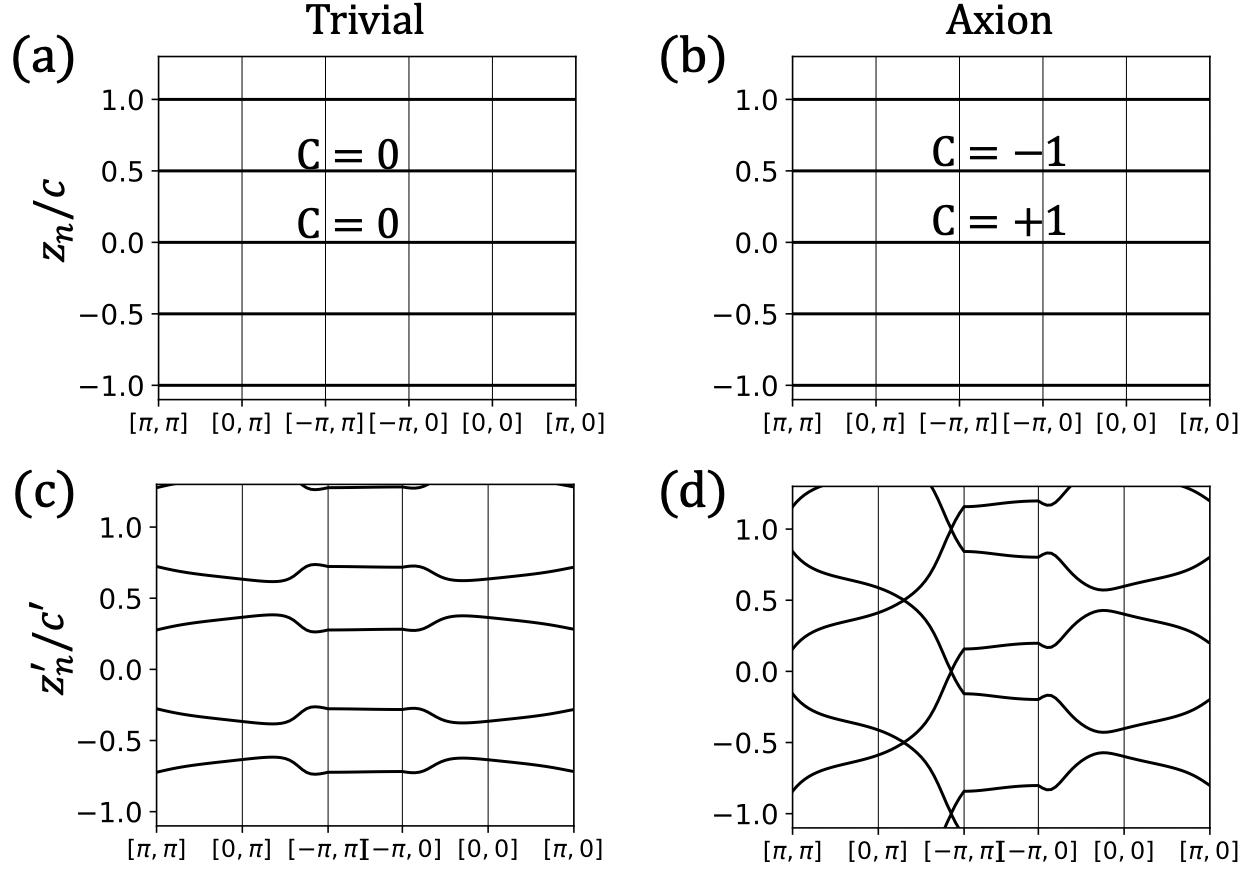}
\caption{Wannier band structures for the two mirror-symmetric phases
  of the alternating Haldane model parametrized by the angle $\phi$ in
  \eq{phi}. (a,c) is the trivial-insulator phase with $\phi=0$, and (b,d) 
  is the axion-insulator phase with
  $\phi=\pi$. The wannierization direction is (001) --~normal to the
  mirror plane~-- in (a,b), and (010) (denoted $\zhat'$)
  --~lying in the mirror plane~-- in (c,d). In
     (a,b), we indicate the Chern numbers of
  the Wannier bands to emphasize that while the Wannier band
  structures look identical, they actually describe different phases.}
\label{fig:mirror}
\end{figure}

We first wannierize the axion-even and odd systems along the direction
(001) normal to the mirror plane, as done in Appendix~A of
Ref.~[\onlinecite{olsen-prb17}].  Within this setting the mirror
operation becomes $M_z$, and the resulting Wannier bands are shown in
the top panels of \fref{mirror}.  The earlier discussion indicates
that a consequence of $M_z$ symmetry is that entire bands can be
pinned exactly at $z\eee0$ or $z\eee c/2$ over the full 2DBZ.  As is
clear from the figure, this is the case here.  In fact, with just two
occupied bands, our model is so simple that this is all we have, with
one Wannier band pinned at $z\eee0$ and the other at $z\eee c/2$.
Just as for the inversion-symmetric case, the axion angle $\theta$ is
$0$ or~$\pi$ depending on whether the Chern index of the
boundary-centered group is even or odd. This is confirmed by
inspection of \fref{cz_alh},
where we show the Chern numbers of the Wannier bands
in a slab geometry.

Finally, we consider the case of wannierization along an axis lying in
the mirror plane, specifically, along $(010)$, which we denote as
$z'$.  Now the mirror operation is of type $M_d$, and the calculated bulk
Wannier bands are displayed in the bottom panels of \fref{mirror}.
\Fref{mirror}(c) shows a disconnected band structure, and
consulting \sref{simple} we conclude that the system is axion-trivial.
Conversely, in the axion-odd phase the Wannier bands are now required
to flow, each with an odd number $\Nd$ of down-touchings. This is
indeed what happens in \fref{mirror}(d), where $\Nd = 1$ for all bands.

\begin{figure}
\centering\includegraphics[width=\columnwidth]{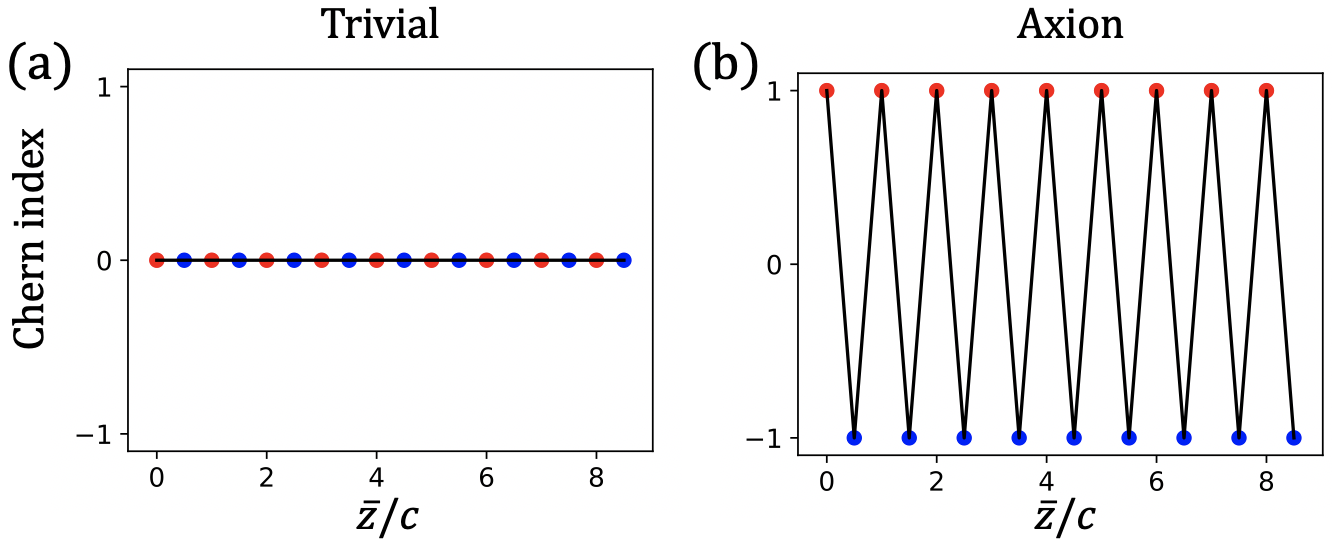}
\caption{Chern index of each isolated slab Wannier band of the
  alternating Haldane model, plotted as a function of its average
  position $\overline{z}$. We consider slabs with a thickness of 10
  unit cells along (001) in (a) the trivial phase of \fref{mirror}(a),
  and (b) the axion phase of \fref{mirror}(b). In the trivial phase
  each Wannier band carries a zero Chern index, while in the
  axion-odd phase the Chern index alternates between $-1$ and
  $+1$.}
\label{fig:cz_alh}
\end{figure}

\subsection{Glide mirror symmetry}
\seclabel{glide}

\subsubsection{General considerations}

A $\zt$ classification of 3D insulators with glide mirror symmetry
was introduced in Refs.~[\onlinecite{fang-prb15}] and
[\onlinecite{shiozaki-prb15}],
and was later argued to be equivalent to the axion $\zt$
classification.\cite{varjas-prb15}
Interestingly, glide mirror symmetry realizes all three types of
axion-odd space-group symmetries, depending on the choice of
wannierization direction.  If $\zhat$ is chosen
as the normal direction to the reflection plane, the glide operation
becomes $\{M_z|\boldsymbol{\tau}_\perp\}$; recalling that our
classification discards
in-plane fractional translations $\boldsymbol{\tau}_\perp$,
here the quantizing symmetry is simply
$M_z$, a $z$-reversing operation, as in \sref{rev}.
Instead, we can choose $\zhat$ to lie in the
reflection plane and be perpendicular to the half-lattice
translation.  In this frame the glide operation is
$\{M_d|\boldsymbol{\tau}_\perp\}$; again discarding
$\boldsymbol{\tau}_\perp$, the axion-odd operation becomes $M_d$, a
symmorphic $z$-preserving operation, per \sref{simple}.
Finally, if $\zhat$ is chosen along the half-lattice translation, then the
glide operation becomes $\ns{M_d}{2}$, a nonsymmorphic $z$-preserving
operation, corresponding to \sref{ns}.

In the numerical tests below we shall consider the latter
nonsymmorphic scenario, which we have not encountered in our previous
examples, as well as the $z$-reversing case.
We will omit the symmorphic $z$-preserving configuration,
since we have already encountered this kind of $M_d$ symmetry in our study
of the alternating Haldane model.

\subsubsection{Model}
\seclabel{glidemodel}

As an example, we take the four-band tight-binding model
at half filling
described in Ref.~[\onlinecite{fang-prb15}] (an equivalent model was
also used in Ref.~[\onlinecite{varjas-prb15}]).  The simple
orthorhombic $a\!\times\!b\!\times\!c$ unit cell
contains two pairs of orbitals with reduced coordinates $(0,0,0)$
and $(0.5,0,0)$, and the Hamiltonian is expressed in $\k$ space as
\beq
\begin{split}
H_{\k} = \sin\left(\frac{k_xa +\phi}{2}\right)\rho_x\tau_x+
\sin(k_yb)\rho_0\tau_y  +
\sin(k_zc)\rho_z\tau_x \\ +
(m - \cos{k_xa} - \cos{k_yb} - \cos{k_zc})\rho_0\tau_z\,.
\eqlab{glidesym}
\end{split}
\eeq
The Pauli matrices $\tau$ can be considered to represent either spin or
orbital degrees of freedom, and the matrices $\rho$ are associated
with sublattice degrees of freedom.

The above model is invariant under the glide operation $\{M_z|a/2\}$
consisting of a reflection about the $z\eee0$ plane followed by a
half-lattice translation along $\xhat$. Because it is so simple, the
model has two additional axion-odd symmetries, $\{M_y|a/2\}$ and
$M_x$, whose presence obscures the role of $\{M_z|a/2\}$ in quantizing
$\theta$. For example, the mirror $M_x$ forces the (100)-oriented
Wannier bands to be completely flat. To break these extra
symmetries, we displace the orbitals to $(0,0,0.1)$ and $(0.5,0,-0.1)$,
and include an additional term
\beq
V_{\k} = 0.4\cos{\frac{k_xa}{2}}\rho_x\tau_z
\eqlab{breaksym}
\eeq
in the Hamiltonian, chosen weak enough so as not to cause any band
inversion.  At half filling, the two lowest bands are filled.  The
model has two adjustable parameters; we set $\phi=0.4$, and vary $m$
to access different phases.

\begin{figure}
\centering\includegraphics[width=\columnwidth]{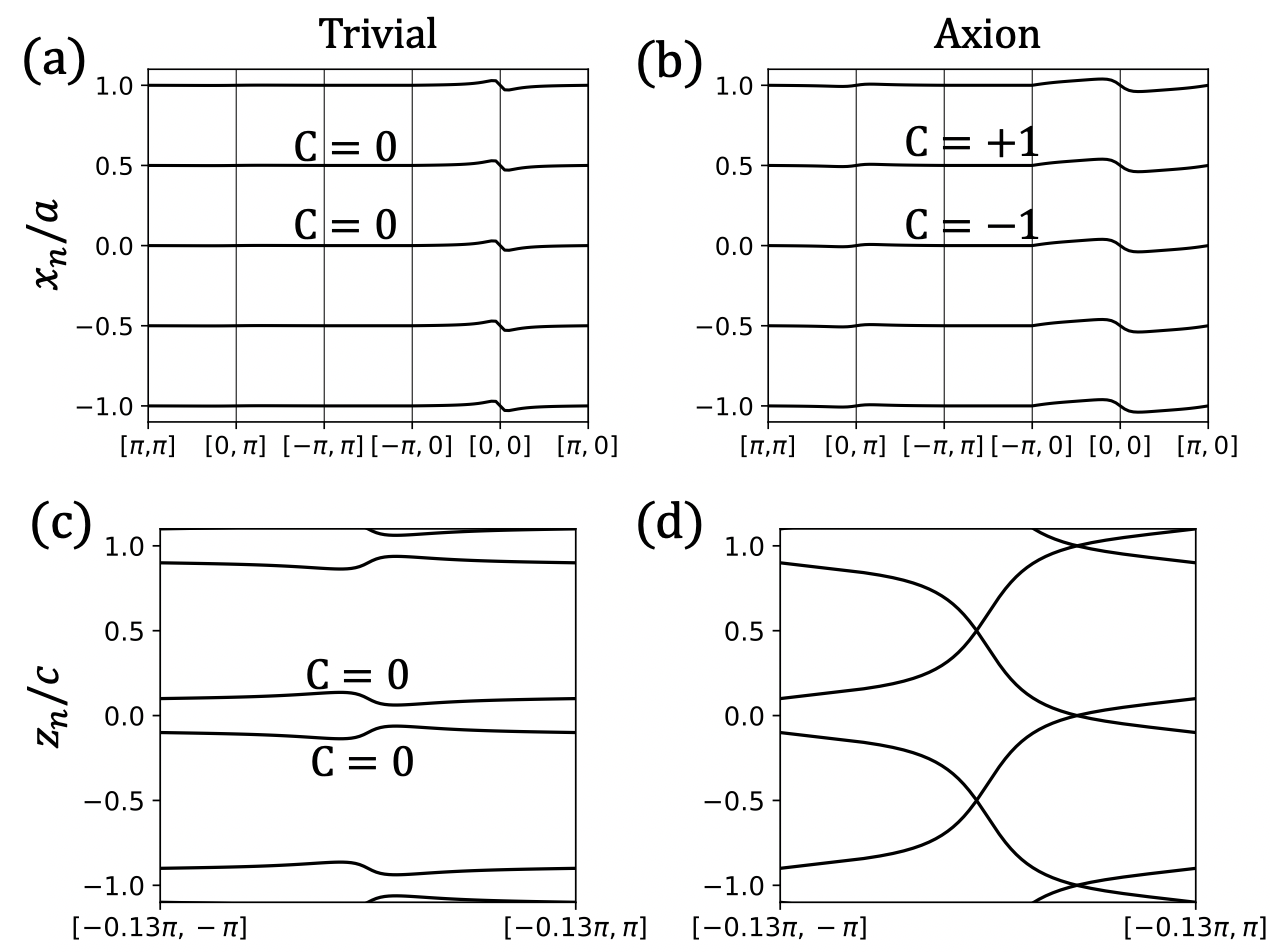}
\caption{Wannier band structures for the model with glide mirror symmetry.
  The trivial and axion phases were obtained by setting
  $(m,\phi) = (3.5,0.4)$ and $(m,\phi) = (2.0,0.4)$ respectively, in
  \eq{glidesym}.  In (a,b) the wannierization direction is
  $\xhat\parallel(100)$, lying in the glide plane and along the
  half-translation; in (c,d) it is $\zhat\parallel(001)$, normal to
  the glide plane. The Chern numbers of isolated bands are indicated
  in panels (a-c).}
\label{fig:glide}
\end{figure}

\Fref{glide} shows the Wannier band structures calculated with $m=3.5$
(trivial phase), and with $m=2.0$ (axion phase), for two different
wannierization directions. In the upper panels the wannierization is
along the half-lattice translation (along $\hat{\bf x}$), and in the
lower panels it is normal to the reflection plane (along
$\hat{\bf z}$).  (For this model, instead of reorienting the axes so
that the wannierization direction is always along $z$ as in the
previous examples, we have chosen to keep the axes fixed.)

In the upper panels of \fref{glide}, the glide operation is
$\{M_d|a/2\}$ in the notation of Table~\ref{tab:g}.
This is a nonsymmorphic operation that preserves the wannierization direction, of the type discussed in \sref{ns}.
In axion-odd phases protected by such symmetries, the Wannier spectrum
is not required to flow. (This is in contrast to axion-odd phases
protected by a simple mirror $M_d$, which must exhibit flow as
illustrated in \fref{mirror}(d) for the alternating Haldane model.)
Indeed, the spectrum of the present model is gapped not only in
the trivial phase of \fref{glide}(a), but also in
the axion-odd phase of \fref{glide}(b).
We can therefore use \eq{CLG} to determine the axion index, taking for
$C_{\rm LG}$ the Chern number of either band.  The Chern numbers of
the two bands are $(0,0)$ in \fref{glide}(a) and $(-1,+1)$ in
\fref{glide}(b), leading to $\theta=0$ and $\theta=\pi$ respectively.

Consider now the lower panels of \fref{glide}, where the glide
operation returns to $\{M_z|a/2\}$
in the notation of Table~\ref{tab:g}.
This is a $z$-reversing operation,
like spatial inversion or a simple mirror $M_z$, so we refer to
\sref{rev}.
Now  the Wannier spectrum is gapped in the trivial phase
of \fref{glide}(c), but connected in the axion phase
of \fref{glide}(d).
(This is different from 
the alternating Haldane model with $M_z$ symmetry, which exhibited
gapped Wannier spectra in both phases.)  In the trivial phase of
\fref{glide}(c), the two isolated bands have vanishing Chern numbers,
so that \eq{half} gives $\theta=0$. For the axion phase of
\fref{glide}(d), which exhibits ``fragile'' flow, we resort to
\eq{evencase}, and since there are single Dirac cones at integer and
half-integer values of $z_n/c$, we conclude that $\theta=\pi$.  
At variance with our previous examples of connected band structures,
here the Dirac nodes are located at generic
low-symmetry points in the 2DBZ, not at high-symmetry points or lines.

\section{Discussion and open questions}
\seclabel{summary}

In this work we have surveyed the symmetry operations that can protect
the axion $\zt$ index in 3D insulators, inducing a
topological classification into those that are trivial (axion-even,
$\theta=0$) and those that are topological (axion-odd,
$\theta=\pi$).
When considered in the context of the hybrid
Wannier construction performed along a given crystallographic
direction, the relevant axion-odd symmetries are classified
into those that involve
reversal along the wannierization direction and those that preserve
it, with the latter case subdivided into those that do not involve
an additional fractional translation along this direction and those
that do.

We then systematically explored the consequences of such
symmetries for the Wannier bands,
and conversely, the means by which the axion $\zt$ index can often be
deduced from characteristics of the Wannier band structure. In
particular, we described the conditions under which
a counting of Dirac touchings between Wannier bands, and/or the
calculation of the Chern number of some central Wannier bands, allows
for a direct determination of the axion index.  We then confirmed our
conclusions and illustrated their consequences for several
paradigmatic cases, namely time reversal, inversion, mirror, and
glide-mirror symmetries.

For some of the cases considered here (including time reversal,
time reversal composed with either a half-lattice
translation\cite{mong-prb10} or a two-fold
rotation,\cite{fang-prb15}
inversion,\cite{hughes-prb11,turner-prb12} and glide-mirror
symmetry\cite{fang-prb15,shiozaki-prb15,varjas-prb15}), the $\zt$
classification has been previously discussed. However, this was
often done on a case-by-case basis by defining the $\zt$ index in a
way that was tailored to each specific symmetry, and was not easily
generalizable to other axion-odd symmetries.
Our work systematizes these considerations,
placing them in the context of a complete classification of axion-odd
symmetries and their influences on Wannier band structures.

Our analysis shows that the Wannier bands are sometimes topologically
required to display ``flow,'' i.e.,
to have touchings between all neighboring bands. Even
when not required, such touchings are often present, usually resulting from
Kramers or symmetry-induced degeneracies at high-symmetry points
in the projected 2DBZ.
In order to carry forward our analysis, we typically assumed that
the only type of touchings between Wannier bands were point Dirac
nodes, i.e., displaying a linear splitting with $\k$ away from the
nodal point.  One interesting avenue for future investigation is to
consider whether there are some symmetries, or combinations of symmetries,
that can give rise to higher-order point nodes or nodal lines
or loops.  In such cases, our prescriptions for deducing the axion
$\zt$ index from the nodal structure of the Wannier bands would need to be
reconsidered.  (An extreme example is the case of a simple
$M_z$ mirror, encountered in \sref{mirror}, which can sometimes produce
two or more entirely flat degenerate Wannier bands.)

In \sref{flow} we also discussed the conditions under which the symmetry
imposes flow on the surface bands of an axion-odd insulator, thus
forcing the surface to be metallic.  This only happens when the
symmetry protecting the bulk axion-odd state is a good symmetry at
the surface, and for some symmetries this is never the case.  For
example, $I$ and $S_4$ symmetries leave no surface facet invariant,
and in these cases all surfaces can generically be gapped. However,
such systems are typically ``second-order topological
insulators,''\cite{benalcazar-sa17,benalcazar-prb17,schindler-sa18,
zhida-prl18,schindler-nat18,langbehn-prb17,ezawa-prb18,khalaf-prb18,
varnava-prb18}
meaning that while the facets themselves are insulating,
the edges where facets meet, or ``hinges,''
are required to be metallic.  This can be understood by recalling that
the axion-odd state with $\theta\eee\pi$ requires all insulating
surfaces to display a half-quantized surface AHC of
$\pm e^2/2h$,\footnote{More generally, the surface AHC could be
   $me^2/2h$ for odd integer $m$.}
and whenever faces with opposite signs of AHC meet, a gap-crossing chiral edge
channel must be present in the hinge band structure, forcing it to be
metallic. 
A more systematic exploration of the relationship of the present
work to the theory of higher-order topological insulators
deserves further attention.

Magnetic axion insulators offer great promise in
spintronic applications,\cite{tokura-nat19} ranging from the
high-temperature quantum anomalous Hall effect to chiral Majorana
fermions.\cite{he-sa17} Unfortunately, and in spite of recent
progress, they still elude experimental observation. Our systematic
approach provides a detailed description of how each axion-odd
symmetry manifests itself in the hybrid Wannier representation. We
have clarified many of the subtle behaviors associated with these
symmetries, and are hopeful that this will facilitate
the discovery of material realizations of novel topological
states exhibiting a quantized CS axion coupling and
half-integer surface quantum anomalous Hall conductivity.
  
\section*{Acknowledgments}

Work by N.V.~and D.V.~was supported by the Institute for Quantum
Matter, an Energy Frontier Research Center funded by the
U.S. Department of Energy, Office of Science, Basic Energy Sciences
under Award No.~DE-SC0019331. Work by I.S. was supported by
Grant No. FIS2016-77188-P from the Spanish Ministerio de Econom\'ia
y Competitividad.

\appendix*

\section{Inversion symmetry and parity counting}
\seclabel{parity-count}

\def\turner{Ref.~[\onlinecite{turner-prb12}]}

\def\btwo{_\textrm{2D}}
\def\bthr{_\textrm{3D}}

\def\zta{Z_2^\mathrm{axi}}
\def\ztp{Z_2^\mathrm{pol}}

\newcommand{\N}{N}  
\newcommand{\M}{M}  
\newcommand{\m}{m}  
\newcommand{\n}{p}  

In an important paper, Turner, Zhang, Mong, and
Vishwanath\cite{turner-prb12} derived a set of rules
for extracting information about the topological state of a 3D
centrosymmetric crystal from a counting of the odd-parity occupied
eigenstates
at the TRIM.  Specifically, they showed that the total number of
odd-parity states summed over the occupied bands at the eight TRIM
must be even in an insulator, and that the axion $\zt$ index
of the system is even or odd if the total number of such
odd-parity states is of the form $4n$ or $4n+2$,
respectively, where $n$ is an integer.
We also call attention to
Refs.~[\onlinecite{hughes-prb11}] and
[\onlinecite{alexandradinata-prb14}] for alternative
viewpoints on the topology of centrosymmetric insulators.

Here we develop a connection
between the numbers of odd-parity Bloch states at the eight TRIM and
the corresponding numbers of odd-parity HW states at the four
PTRIM, and use this to develop criteria for deciding
the axion $\zt$ index based on an inspection of the Wannier band
structure. The results are consistent with those of \sref{I}, thus
providing an alternative derivation of the results presented there.

As a preliminary exercise, we begin by providing a pedagogical
development of some important rules that underlie the theory of
\turner, beginning with a 1D centrosymmetric insulator and
working our way to two and three dimensions.

\subsection{Parity counting for 1D insulators}
\seclabel{oned}

Consider a 1D centrosymmetric insulator of lattice constant $c$,
with its two inversion centers located at the origin $z\eee0$,
denoted as location $A$, and at $z\eee c/2$, denoted as $B$.
In reciprocal space, there are $J$ occupied bands at each of the
two TRIM at $\Gamma$ ($k\eee0$) and $X$ ($k\eee\pi$; reduced
wavevectors are used here). Let $\N_\mu\oo$ be the number of
odd-parity Bloch eigenstates at $\mu\eee\Gamma$ or $\mu\eee X$, with
$\N\oo=\sum_\mu \N_\mu\oo$ summed over $\Gamma$ and $X$,
where the parities of the Bloch states are defined
relative to the chosen origin at $A$.

Now consider the maximally localized Wannier centers for this 1D crystal
and the corresponding Wannier functions.  In view of the inversion
symmetry, the general description is that we find
$\M_A\oo$ and $\M_A\ee$ Wannier states centered at $A$ with odd and
even parity, respectively, about $A$;
$\M_B\oo$ and $\M_B\ee$ states centered at $B$ with odd and even
parity, respectively, about $B$;
and $\M_P$ pairs of Wannier states centered at
$\pm z$, distinct from both $A$ and $B$.  The total number
of occupied states is thus
\beq
J=\M_A\oo+\M_A\ee+\M_B\oo+\M_B\ee+2\M_P \,.
\eqlab{Jsum}
\eeq
We can also write down the total number $\N\oo$ of odd-parity
Bloch states summed over the two TRIM as
\beq
\N\oo=2\M_A\oo+\M_B\oo+\M_B\ee+2\M_P \,.
\eqlab{modd}
\eeq
This follows because
an odd-parity Wannier state at $A$ contributes odd-parity Bloch states at
both $\Gamma$ and $X$;
an even-parity one at $A$ contributes none at $\Gamma$ or $X$;
an odd-parity one at $B$ contributes an odd-parity state only at $\Gamma$;
an even-parity one at $B$ contributes an odd-parity state only at $X$;
and the pair contributes one odd-parity state at each of $\Gamma$ and $X$.
From \eqr{Jsum}{modd} it follows trivially that
\beq
\N\oo=J+\M_A\oo-\M_A\ee \,.
\eqlab{Jplus}
\eeq

These considerations provide an elementary derivation of a 1D
polarization-parity relation.  A 1D centrosymmetric insulator has
a polarization $\zt$ index,
which we shall refer to as $\ztp$, defined as $0$ or $1$ if the total
Berry phase of the occupied Bloch bands is $0$ or $\pi$, or the
electronic contribution to the polarization is $0$ or $e/2$,
respectively.  Each Wannier center at location $B$ contributes one
unit to $\ztp$, while those at $A$ do not contribute, and
neither do the pairs.  Thus, $\ztp=\M_B\oo+\M_B\ee$ mod 2.
But this is equal to \eq{modd} mod 2, so it follows that
\beq
\ztp = \N\oo \quad\hbox{mod 2} \,.
\eqlab{1drel}
\eeq
That is, the Bloch parity sum determines whether the 1D electronic
polarization is trivial (Wannier charge is centered at $A$)
or not (centered at $B$).  Put another way, the
polarization $\zt$ index and the parity $\zt$ index are the same in
1D.

\subsection{Parity counting for 2D insulators}
\seclabel{twods}

The authors of \turner\ also derived a Chern-parity relation for a 2D
insulator. Let $N\oo\btwo$ be the number of
odd-parity occupied Bloch eigenstates obtained by summing over the
four TRIM of the 2DBZ.  They showed that
\beq
\N\oo\btwo = C \quad\hbox{mod 2}
\eqlab{2drel}
\eeq
where $C$ is the total Chern number of the 2D insulator (again summed
over all occupied bands).  This follows from the considerations of the
previous subsection.  Let the system extend in the $y$-$z$ plane, and
we wannierize along $z$ as a function of $k_y$.  The polarization
$P_z(k_y)$ is well defined and varies smoothly for intermediate $k_y$,
only taking quantized values  $\{0,e/2\}$ at
$0$ and $\pi$. As is well known, the Chern number $C$ describes the
pumping of polarization $P_z$ as $k_y$ is taken adiabatically from $0$
to $2\pi$. A change of  $P_z$ by $-e$,
or equivalently a change of the Berry phase by $2\pi$, corresponds to
a unit Chern number. In view of the inversion symmetry, half of the
pumping of  $P_z$ occurs as $k_y$ traverses from 0
to $\pi$, and the other half
between $\pi$ and $2\pi$.  Thus, the value of $\ztp$
changes from $k_y\eee0$ to $k_y\eee\pi$ if and only
if the Chern number is odd.

On the other hand, the earlier 1D analysis shows that the
value of $\ztp$ at $k_y\eee0$ ($k_y\eee\pi$) corresponds to
the number of odd-parity occupied eigenstates summed over the two TRIM
projecting onto $k_y\eee0$ ($k_y\eee\pi$).
Thus, it follows that the value of $\ztp$
changes from $k_y\eee0$ to $k_y\eee\pi$, and hence the
Chern number is odd, if and only if $\N\oo\btwo$, the sum of odd
Bloch parities over all four TRIM, is odd. This proves \eq{2drel}.

Our considerations will be limited to the case of insulators
with $C\eee0$, in which case \eq{2drel} implies that $\N\oo\btwo$
must be even.  This fact will be useful later.

\subsection{Parity counting for 3D insulators}
\seclabel{threed}

We now apply  the rules derived above to the consideration of
a 3D centrosymmetric insulating crystal.
Let $C_z$ be the Chern index on the plane $k_z\eee0$ or
$k_z\eee\pi$; these are equal since for an insulator $C_z$
must be the same at at all $k_z$. Now $\N\oo\bthr$, the total
number of odd-parity Bloch states at the eight TRIM, is
just the sum of those at $k_z\eee0$ and $k_z\eee\pi$, and
using \eq{2drel}, we have that
\beq
\N\oo\bthr = 2C_z = 0 \quad\hbox{mod 2} \,.
\eqlab{3da}
\eeq
This provides a rederivation of one of the important results given in
\turner. As those authors pointed out, this implies that any system
with $\N\oo\bthr$ odd cannot be insulating; it must have nodes of
degeneracy between the nominal valence and conduction bands, as in a
Weyl semimetal. (In this work we assume vanishing Chern numbers, but
the above result is general.)

The authors of \turner\ derived another result that is central here,
namely that the axion $\zt$ index, which we now denote as
$Z_2^\textrm{axi}$ in analogy with $Z_2^\textrm{pol}$ in 1D, is
given by
\beq
Z_2^\textrm{axi} = \N\oo\bthr/2\quad\hbox{mod 2}\,.
\eqlab{3db}
\eeq
In other words, if $\N\oo\bthr$ is of the form $4n$ for integer $n$,
the system is axion-even, while if $\N\oo\bthr=4n+2$, it is
axion-odd.  This is a rather deep result which we shall not
attempt to rederive here.  Our goal now is to use this result
to determine the axion $\zt$ index from an inspection of the
Wannier band structure.

We choose to wannierize along $z$ and retain the $(k_x,k_y)$ Bloch
labels.  Let $\mu$ label the four PTRIM
in the 2DBZ in $(k_x,k_y)$ space; each of these corresponds to a pair
of the 3D TRIM separated by $\pi$ along $k_z$.  For a given PTRIM
$\mu$, the string of $\k$ points extending from $k_z\eee0$ to $2\pi$
describes a 1D centrosymmetric insulator.  We can then apply the
analysis of \sref{oned}, rewriting \eq{Jplus} as
\beq
\N_\mu\oo=J+\M\bmA\oo-\M\bmA\ee \,,
\eqlab{Jpmu}
\eeq
where $\N_\mu\oo$ is the total number of odd-parity Bloch states
at the two associated TRIM.
This motivates us to define
\beq
\m_A=\sum_{\mu=1}^4 (\M\bmA\oo-\M\bmA\ee)
\eqlab{mAdef}
\eeq
as the excess number of odd over even-parity
HW states pinned at location $A$, summed over the four PTRIM.
Then, noting that $J$ is an integer, \eq{Jpmu} implies that
\beq
\N\oo\bthr=\sum_{\mu=1}^{4} 
\N_\mu\oo
=\m_A \quad\hbox{mod 4} \,.
\eqlab{Moosum}
\eeq
Because of \eq{3da}, we know that a 3D insulator must have an even
value of $\N\oo\bthr$, so $\m_A/2$ is an
integer.  Combining \eq{Moosum} with \eq{3db}, we arrive at the first
major result of this Appendix, namely that the axion $\zt$ index is
simply given by
\beq
Z_2^\textrm{axi} = \m_A/2 \quad\hbox{mod 2}\,.
\eqlab{zaxp}
\eeq
That is, the system is axion-odd if and only
if $\m_A/2$ is an odd integer.

Again we point out that the choice of origin at $A$ was arbitrary;
placing the origin at $B$ leads to the conclusion that the axion
$\zt$ index is also determined by $\m_B=\sum_{\mu=1}^4 (\M\bmB\oo-\M\bmB\ee)$,
which must equal $m_A$ mod 4.

At one level, \eq{zaxp} provides an answer to the question of how
to extract the axion $\zt$ index within the HW representation.
One simply looks for Wannier bands pinned at $z\eee0$ (or $z\eee c/2$)
at the PTRIM, computes the parities of the corresponding HW states,
and sums the parities over the four PTRIM.
Half this number gives $\zta$.  Unfortunately, this approach does
require an extraction of the parities of these HW states, so the
needed information does not appear to be contained in the Wannier band
structure itself.  However, we show next that this is not always
true; a simple count of the number of such pinned states,
without a knowledge of their parities, is sometime sufficient.

\subsection{Node counting for 3D insulators}
\seclabel{nodes}

Again we restrict ourselves to 3D centrosymmetric insulators with
all zero Chern indices.  It is useful to repeat the definition of
\eq{mAdef} for each PTRIM individually, and also to define the
corresponding count $\n\bmA$ of the total number of Wannier bands pinned
at $A$, via
\bea
\m\bmA&=&\M\bmA\oo-\M\bmA\ee \,,
\eqlab{mmdef}
\\
\n\bmA&=&\M\bmA\oo+\M\bmA\ee \,.
\eqlab{mndef}
\eea
With this notation, \eq{Jpmu} becomes
\beq
\N_\mu\oo=J+\m\bmA\,.
\eqlab{Jpmu-b}
\eeq

Now consider two neighboring PTRIM $\mu$ and $\mu'$; these define a 2D
centrosymmetric system corresponding to the $k$-space plane in 3D
passing through the four TRIM projecting onto these two PTRIM.
Following the discussion at the end of \sref{twods} of this Appendix,
the zero-Chern assumption requires that the total number of
odd-parity Bloch states on these four TRIM, i.e.,
$\N_\mu\oo+\N_{\mu'}\oo$, must be even.
Summing \eq{Jpmu-b} over $\mu$ and $\mu'$ and noting that $J$ is
an integer, it follows that $\m\bmA+\m_{\mu'A}$ is even.
But \eqs{mmdef}{mndef} differ by an even integer, so
$\n\bmA+\n_{\mu'A}$ is also even.  That is, $\n\bmA$ and
$\n_{\mu'A}$ are either both even or both be odd. Continuing to
apply this principle to other pairs of PTRIM, we come to an important
constraint, namely that the number $\n\bmA$ of Wannier bands pinned at $A$
is either even at all four PTRIM, or odd at all four PTRIM.  We
consider these two cases in turn.

We first assume that $\n\bmA$
is even at all four PTRIM.  In order to make further progress, we
assume henceforth that at any given PTRIM, the HW states pinned at $A$
are all of the same parity.  We call this the ``uniform parity''
assumption.  This is a rather natural assumption, since states of even
and odd parity about $A$ will generically have a nonzero $z$ matrix
element between them.  The maximal localization procedure corresponds
to diagonalizing the $z$ operator in the occupied band
space,\cite{kivelson-prb82,marzari-prb97}
and this will generically result in hybridization and splitting to
form a pair of HW states at $\pm z$, so that these
states will no longer be pinned at
$A$. If this process repeats until the only states left at $A$ are all
of the same parity, then our uniform parity assumption holds.%
\footnote{Note that the Wannier bands correspond to the spectrum of the
  projected $z$ operator, which anticommutes with inversion.
  Thus, the projected $z$ operator acts like a sublattice
  (chiral) symmetry.  Lieb's theorem\cite{lieb-prl89} states that if
  there is an excess of $N$ even- or odd-parity states in a
  sublattice-symmetry system, there must be at least $N$ eigenvalues
  pinned at $z\eee0$.  Here we are assuming the minimum, i.e., that
  only $N$ eigenvalues remain at $z\eee0$.}
This procedure may break down if symmetries other than inversion are
present (e.g., the even and odd parity states at one of the TRIM fail
to split because they belong to different irreps under rotation about
the $z$ axis), or in case the system has been fine-tuned to force the
splitting to vanish.  Therefore, we treat it simply as an assumption
in the following.

Now with this assumption, $|\m\bmA|=\n\bmA$ and both are even,
so $\m\bmA=\n\bmA$ mod 4, and defining $\n_A=\sum_{\mu=1}^4 \n\bmA$,
\eq{zaxp} can be replaced by
\beq
Z_2^\textrm{axi} = \n_A/2 \quad\hbox{mod 2}\,.
\eqlab{zaxpp}
\eeq
Thus, the axion $\zt$ index is obtained just by
counting the number of HW states at the PTRIM that are
pinned at $A$!  Again, the choice of origin at $A$ was arbitrary, so
the same applies to the counting of HW states pinned at $B$.  Assuming
these PTRIM degeneracies are point nodes of contact between Wannier bands,
$\n_A/2$ is just the number of Dirac crossings at the PTRIM at $A$, and
the system is axion-odd if and only if this number of Dirac crossings
is odd.  The same applies to the Dirac nodes at $B$.  This is in
perfect agreement with the conclusions of \sref{rev} as expressed in
\eq{evencase} and the discussion following it.

In case the number $\n\bmA$ of Wannier bands pinned at $A$ is odd at each
of the PTRIM, and again assuming uniform parity of the HW states
at each of these four points individually, we can use a similar
analysis to evaluate the contribution from all except the central
Wannier band; the contribution of these is again given by \eq{zaxpp}.
To obtain the contribution of the central Wannier band, we need to
know the sum of the parities over the four PTRIM for this band.
If we have access to the parities, we can do this directly.
Alternatively, a computation of the Chern number of
this band will also suffice, since \eq{2drel} tells us that
the Chern index is odd if and only if the parity sum is odd.
In this way we again arrive at \eq{oddcase}, which involves the
computation one Chern number in addition to a counting of the
Dirac cones passing through $z\eee0$.

Once again we note that the same analysis can equally well be applied
at location $B$, and the choice of inversion center to use for the
analysis can be made on the basis of convenience.  In any case, we
have succeeded in arriving at the same conclusions as were presented
in the main text at the end of \sref{rev}, but from a very different
point of view.

\bibliography{pap}
\end{document}